\documentclass[12pt]{article}
\usepackage{amssymb}
\usepackage{amsmath}
\usepackage{youngtab}
\usepackage{cite}
\usepackage[]{hyperref}
\usepackage{multirow}
\usepackage{MnSymbol}
\usepackage{bbm}
\usepackage{dcolumn}
\usepackage{tikz}
\input xy
\xyoption{all}

\numberwithin{equation}{section}

\newcommand{\be}{\begin{equation}}
\newcommand{\ee}{\end{equation}}
\newcommand{\bea}{\begin{eqnarray}}
\newcommand{\eea}{\end{eqnarray}}

\begin{document}

\begin{center}

{\large\bf GLSMs for exotic Grassmannians}

\vspace{0.2in}

Wei Gu$^1$, Eric Sharpe$^2$, Hao Zou$^2$

\begin{tabular}{cc}
{\begin{tabular}{l}
$^1$ Center for Mathematical Sciences\\
Harvard University\\
Cambridge, MA  02138
\end{tabular}} &
{\begin{tabular}{l}
$^2$ Dep't of Physics\\
Virginia Tech\\
850 West Campus Dr.\\
Blacksburg, VA  24061
\end{tabular}}
\end{tabular}

{\tt weigu@cmsa.fas.harvard.edu},
{\tt ersharpe@vt.edu},
{\tt hzou@vt.edu}

\end{center}

In this paper we explore nonabelian gauged linear sigma models (GLSMs) 
for symplectic and orthogonal
Grassmannians and flag manifolds, 
checking e.g. global symmetries, Witten indices, and Calabi-Yau conditions,
following up a proposal in the math community.  
For symplectic Grassmannians, we check that Coulomb branch
vacua of the GLSM
are consistent with ordinary and equivariant quantum cohomology of the space.

\begin{flushleft}
August 2020
\end{flushleft}

\newpage
\tableofcontents

\newpage

\section{Introduction}

Gauged linear sigma models (GLSMs) \cite{Witten:1993yc}
have proven to be extraordinary
physical tools to examine a wide range of questions in string
theory and string compactifications, ranging from global properties
of moduli spaces of SCFTs for Calabi-Yau compactifications to
representations of quantum cohomology rings.  The bulk of that work
has focused on abelian two-dimensional theories, but in recent years
technology has developed to the point where we can make inroads on
understanding nonabelian theories.

To further that program, one of the tasks one must accomplish is to find
physical descriptions of more geometries.  For example, one can
write down nonabelian GLSMs which have nontrivial IR fixed points,
but to efficiently compute e.g. chiral rings, it helps enormously if one
can interpret the resulting phases geometrically.  To this end, in this paper
we will explore (nonabelian) GLSMs for some additional spaces,
namely symplectic and orthogonal Grassmannians, following up
a brief proposal in \cite{ot}.  We will check that description by
e.g. comparing physically-derived quantum cohomology rings against
known mathematics results, and study
the phases and other properties of the GLSMs.

Ordinary Grassmannians $G(k,n)$ can be described with
GLSMs
using methods that have been known for
a long time, going back to \cite{Witten:1993xi}.
They have played an important role in many papers.
However, they are not the only notion of Grassmannians known to
mathematicians.
There are other Grassmannians in the mathematics literature,
notably the symplectic and orthogonal Grassmannians.
These also occasionally arise in physics,
see e.g. \cite{Eager:2019zrc} and references therein,
but aside from a brief proposal in \cite{ot}, their GLSM realizations
have not been studied at all.
The purpose of this
paper is to fill this gap, following up the proposal of \cite{ot}
by comparing GLSM predictions for quantum cohomology rings, Witten
indices, Calabi-Yau conditions, and studying the GLSM phases.

Possible Grassmannians and flag manifolds\footnote{
In most of this paper, for simplicity
we focus on Grassmannians, but analogues for
flag manifolds do exist, and we discuss corresponding GLSMs later
in this paper.
}
are given mathematically as cosets $G/P$,
with $P$ a parabolic subgroup of $G$, and $G$ describing the
symmetries of the space, which also correspond to
global symmetries of the corresponding physical theory.  
We list below some examples from
\cite[section 23.3]{fulton}:
\begin{itemize}
\item $A_n$:  These are the Grassmannians $G(k,n+1) = SL(n+1)/P$, 
which have global
symmetry
\begin{equation}
\frac{U(n+1)}{U(1)} \: = \: P SU(n+1) \: = \:
\frac{ SU(n+1) }{ {\mathbb Z}_{n+1} }.
\end{equation}
\item $B_n$:  These are the orthogonal Grassmannians $OG(n,2n+1) = SO(2n+1,{\mathbb C})/P$.
\item $C_n$:  These are the symplectic and 
Lagrangian Grassmannians $SG(k,2n), LG(n,2n) = Sp(2n,{\mathbb C})/P$.
\item $D_n$:  These are the orthogonal Grassmannians $OG(n,2n) = SO(2n,{\mathbb C})/P$.
\end{itemize}
(The various Grassmannians above are sometimes referred to as type $A$,
$B$, $C$, $D$ Grassmannians respectively, in reference to their 
symmetries.)  
In each of these cases, the global symmetry group of the GLSM for $G/P$ is
given by $G$ (up to finite quotients), and has Lie algebra indicated by
the classification above.  
(This is simply the subgroup of the group $PSU(2n)$ or
$PSU(2n+1)$ of
rotations of the chiral primaries that preserves the superpotential,
which requires that either a metric or symplectic form be preserved.)
For example, for the $B_n$ series, we will see the GLSM has $2n+1$ chiral
primaries, and a superpotential defined by a metric on those chiral
primaries.  The resulting symmetry group is the subgroup of
$PSU(2n+1)$ that preserves that metric -- hence, some finite group
quotient of $SO(2n+1)$, corresponding to $B_n$.

In addition to the $A$, $B$, $C$, and $D$ type Grassmannians and flag
manifolds, one can also obtain Grassmannians and flag manifolds from
exceptional groups.  We will leave the development of their GLSMs to
future work.

Let us also remark on mirrors.  In mathematics, there is a notion of
mirrors to homogeneous spaces, see e.g. \cite{r08,mr,rw,pr}.
When we speak about mirrors to nonabelian theories, we will be using
a slightly different mirror symmetry construction, described in
\cite{Gu:2018fpm}.  
These two constructions were compared in \cite{Gu:2018fpm}; briefly,
although in general they give different Landau-Ginzburg models, all we
really are concerned with in a Landau-Ginzburg model is its IR behavior,
encoded in its critical loci, and at least on the face of it,
these different-looking constructions of Landau-Ginzburg mirrors seem to
encode the same IR physics, at least so far as we are aware.

We begin in section~\ref{sec:lag} by describing GLSMs for symplectic
Grassmannians $SG(k,2n)$ and flag manifolds, the type $C$ spaces listed
above.  These can be understood as submanifolds of ordinary Grassmannians
$G(k,2n)$ and flag manifolds satisfying an isotropy condition, which is the
key to the GLSM we present.  We check our description by comparing ordinary
and equivariant quantum cohomology rings arising in the GLSM to those
arising in mathematics, as well as by comparing Witten indices across
different phases.  We also check that the Calabi-Yau condition arising
physically matches that in mathematics.  Finally, we discuss mirrors of
the GLSMs for symplectic Grassmannians.

In section~\ref{sect:og} we perform the analogous analyses for
orthogonal Grassmannians $OG(k,n)$ and flag manifolds, the type $B$ and $D$
spaces listed above.  After proposing
GLSMs for these spaces, we study
the mixed Higgs-Coulomb phases arising for $r \ll 0$, and compare the
Calabi-Yau condition arising physically in these GLSMs to that arising
mathematically.  Finally, we discuss mirrors of the GLSMs for orthogonal
Grassmannians.

In several appendices we collect various technical computations which
supplement and clarify the computations in the text.

To be clear, our paper is not the first to describe GLSMs for all of these
cases: GLSMs for symplectic and orthogonal Grassmannians
were proposed at the end of \cite{ot}.  The purpose of this paper is to
more systematically analyze the physics of
these theories, carefully checking that the GLSMs
have the desired properties (quantum cohomology relations, Witten indices,
global symmetries,
and so forth), as well as to explore novel physical aspects of these
theories.

\section{Symplectic Grassmannians \texorpdfstring{$SG(k,2n)$}{SG(k,2n)}}
\label{sec:lag}

\subsection{Background and GLSM realization}
\label{sect:sg:background}

The symplectic Grassmannian $SG(k,2n)$ is a space 
parameterizing $k$-dimensional subspaces of ${\mathbb C}^{2n}$ 
which are isotropic with respect to a symplectic form on
${\mathbb C}^{2n}$.
This can be described more explicitly as a subvariety of an 
ordinary Grassmannian $G(k,2n)$, a fact that will be used in GLSM
realizations.
The dimension of $SG(k,2n)$ is given by
by \cite{coskun}, \cite[section 3.1]{sam-weyman}
\begin{equation}
2nk - \frac{k(3k-1)}{2}.
\end{equation}
When $k=n$, $SG(n,2n)$ is the space of maximal isotropic subspaces of 
${\mathbb C}^{2n}$, which is also known
as the  Lagrangian Grassmannian, and often denoted $LG(n,2n)$. 
In the case $k=1$, the isotropy condition trivializes, and \cite{coskun}
\begin{equation}
SG(1,2n) \: \cong \: G(1,2n) \: \cong \: {\mathbb P}^{2n-1},
\end{equation}
as we shall see explicitly in GLSMs in a moment.
Another common special case is
$SG(2,4) = LG(2,4) \cong {\mathbb P}^4[2]$, as is mentioned in
appendix~\ref{app:dualities}.

The Euler characteristic of $SG(k,2n)$, relevant for Witten index
computations, is given by \cite{wli}
\begin{equation}
2^k \left( \begin{array}{c} n \\ k \end{array} \right).
\end{equation}
In particular, the Euler characteristic of the Lagrangian
Grassmannian $LG(n,2n)$ is $2^n$.
Some background on symplectic Grassmannians can be found in {\it e.g.}
\cite{coskun,sam-weyman}, \cite[section 6.3]{araujo}.

The GLSM for a symplectic Grassmannian implements an isotropy condition
on top of an ordinary Grassmannian, so 
let us first
quickly review the GLSM for an ordinary Grassmannian $G(k,N)$,
following \cite{Witten:1993xi}.  This is a $U(k)$ gauge theory with
$N$ chirals in the fundamental representation, and no superpotential.
The D-terms of the theory can be interpreted as the statement that the
$N$ fundamental-valued chirals form a set of $k$ orthogonal, normalized
vectors in ${\mathbb C}^N$, and gauging the $U(k)$ effectively quotients out
the rotations, hence this describes $k$-dimensional subspaces of 
${\mathbb C}^N$.  The GLSM for a symplectic Grassmannian implements an
isotropy condition on top of an ordinary Grassmannian.

The GLSM for the symplectic Grassmannian $SG(k,2n)$ 
is then a $U(k)$ gauge theory with $2n$ chirals $\Phi^a_{\pm i}$
in the fundamental representation $V$ ($a \in \{1, \cdots, k\}$,
$i \in \{\pm 1, \pm 2, \cdots, \pm n \}$), and one chiral superfield
$q_{ab}$ in the representation $\wedge^2 V^*$, with superpotential
\begin{equation} \label{eq:lagsuperpotential} 
W 
\: = \: \sum_{\alpha, \beta} q_{ab} \Phi^a_{\alpha} \Phi^b_{\beta} 
\, \omega^{\alpha \beta} 
\: = \: \sum_{i=1}^n q_{ab} \Phi^a_i \Phi^b_{-i}.
\end{equation}
The superpotential realizes an isotropy condition with respect to
a symplectic form 
\begin{equation}
\omega \: = \:
\left[ \begin{array}{cc}
0 & I_n \\
- I_n & 0 \end{array} \right]
\end{equation}
on ${\mathbb C}^{2n}$, following from the fact that
\begin{equation}
\left[ \phi^a_i, \phi^a_{-i} \right]
\left[ \begin{array}{cc}
0 & I_n \\
- I_n & 0 \end{array} \right]
\left[ \begin{array}{c}
\phi^b_i \\ \phi^b_{-i} \end{array} \right] \: = \:
\sum_i \left( \phi^a_i \phi^b_{-i} -
\phi^a_{-i} \phi^b_i \right).
\end{equation}
(These GLSMs were also briefly described in
\cite[section 2.4.3, example 3]{ot}.)

The isotropy condition resulting from the superpotential above takes
the simple form
\begin{equation} \label{eq:sgisotropy}
\sum_i \phi^a_i \phi^b_{-i} \: = \:
\sum_i \phi^b_i \phi^a_{-i}.
\end{equation}
In the special case $k=1$, this condition is satisfied trivially.
In this special case, there are no
$q_{ab}$ (since $\wedge^2 {\bf 1}$ is empty) and the superpotential
is not present.  As a result, $SG(1,2n)$ coincides with the ambient
$G(1,2n) = {\mathbb P}^{2n-1}$.

The global symmetries of this theory are rotations of the chiral
superfields that are compatible with the superpotential.
Specifically, rotations of the chiral superfields themselves are
represented by the group
\begin{equation}
U(2n)/U(1) \: = \: PSU(2n) \: = \: SU(2n)/{\mathbb Z}_{2n}.
\end{equation}
The rotations that preserve the superpotential are precisely those
which preserve the symplectic form, hence the global symmetry
group is $Sp(2n,{\mathbb C})$.

Clearly, symplectic Grassmannians $SG(k,2n)$ can be embedded into ordinary
Grassmannians $G(k,2n)$.  The Pl\"ucker embedding of ordinary Grassmannians,
realized physically as $SU(k)$-invariant baryons
\begin{equation}
B_{\alpha_1 \cdots \alpha_k} \: = \:
\epsilon_{a_1 \cdots a_k} \phi^{a_1}_{\alpha_1} \cdots \phi^{a_k}_{\alpha_k},
\end{equation}
for $\alpha = \pm i$, also is relevant for symplectic Grassmannians.
Just as for ordinary Grassmannians, these define a map into a projective
space of dimension
\begin{equation}
\left( \begin{array}{c} 2n \\ k \end{array} \right) \: - \: 1.
\end{equation}
In the special case that $k=n$, there is a second class of $SU(k)$-invariant
operators, that also define a map.  These operators are given by
\begin{equation}
\tilde{B}_{\pm 1, \cdots, \pm n} \: = \:
\epsilon_{a_1 \cdots a_n} \phi^{a_1}_{\pm 1} \cdots \phi^{a_n}_{\pm n},
\end{equation}
and they define a map from $LG(n,2n)$ into a projective space of dimension
$2^n - 1$.  (See \cite[section 3.3]{gm} for more information.)

For later use, since one of the matter representations is slightly
unusual, we give here the $D$-terms:
\begin{equation}
\frac{1}{e^2} D_a^b \: = \: \sum_{i=1}^n\left(\overline{\phi}_a^i\phi^b_i + \overline{\phi}_a^{-i}\phi^b_{-i}\right) - 2 \overline{q}^{bc}q_{ac} - r\delta_a^b.
\end{equation}

In the special case that $k$ is odd,
the GLSM for $SG(k,2n)$ has no Higgs branch for $r \ll 0$.
This follows from the diagonal terms in the $D$ terms above,
and the fact that since $q$ is an antisymmetric matrix,
it can only have an even number of eigenvalues.  The matrix $q$
can therefore be put in the form
\begin{equation}
\left[ \begin{array}{cccc}
0 & * & \cdots & 0 \\
-* & 0 & \cdots & 0 \\
\vdots & \vdots & \vdots & 0 \\
0 & 0 & \cdots & 0 \end{array} \right],
\end{equation}
with one vanishing row and one vanishing column.
In this basis, the diagonal $D$ term corresponding to the bottom right entry
has the form
\begin{eqnarray}
        \frac{1}{e^2} D_a^a & = & \sum_{i=1}^n\left(\overline{\phi}_a^i\phi^a_i + \overline{\phi}_a^{-i}\phi^a_{-i}\right) - r,
\\
& = &
\sum_{i=1}^n \left( \left| \phi_i^a \right|^2 + 
\left| \phi_{-i}^a \right|^2 \right) - r,
\end{eqnarray}
with no sum over $a$ and no $q$ fields.
This $D$ term has no vanishing solutions for $r \ll 0$, hence there can be
no Higgs phase for $r \ll 0$ if $k$ is odd.

The cases of $k$ even are different.  Here, there does appear to be
a Landau-Ginzburg phase, which contributes to the Witten index.
Consider the case $SG(2,2n)$.
Here, there is one $q$ field, $q_{12}$, which transforms only
under the overall $U(1) = \det U(2)$.  In the $r \ll 0$ phase,
it Higgses the gauge group $U(2)$ to $SU(2) \subset U(2)$.
From \cite[section 3.2]{Hori:2006dk}, \cite[section 4.7]{Benini:2013xpa},
the Witten index of an
$SU(2)$ gauge theory with $N$ fundamentals is
\begin{equation}
\left\{ \begin{array}{cl}
(1/2)(N-1) & N \, {\rm odd}, \\
(1/2)(N-2) & N \, {\rm even},
\end{array}
\right.
\end{equation}
so we see that the Higgs branch of the $r \ll 0$ phase of the
$SG(2,2n)$ GLSM has Witten index 
\begin{equation}
(1/2)(2n-2) \: = \: n-1.
\end{equation}
For example, the $r \ll 0$ phase of the GLSM for $LG(2,4)$ has
Witten index $2-1 = 1$.

For $SG(k,2n)$ for even $k > 2$, we expect a similar story.  
For $r \ll 0$, the $q$ fields get a vev, which Higgses\footnote{
As a quick consistency check, note that
the difference in dimensions of the gauge
groups
\begin{displaymath}
\dim U(k) - \dim Sp(k) \: = \: k^2 - (1/2) k (k+1) \: = \: (1/2) k (k-1)
\end{displaymath}
matches the dimension of the representation $\wedge^2 {\bf k}$.
} $U(k)$ to
$Sp(k)$ (see e.g. \cite{haber}).
Beyond that, we do not have a complete understanding.  Based on the
fact that there are $k(k-1)$ off-diagonal D terms generating relations,
we (naively) suspect that this theory has the same Witten index as a 
two-dimension $Sp(k)$ gauge theory with $2nk - k(k-1)$ chirals, in
$2n-k+1$ copies of the fundamental representation, which from
\cite[equ'n (5.10)]{Hori:2011pd} we expect should have Witten index
\begin{equation}
\left( \begin{array}{c} (1/2)(2n-k) \\ k/2 \end{array} \right).
\end{equation}
We will check this conjecture numerically in a pair of examples
in table~\ref{table:witten-check} in section~\ref{sect:sg:witten},
and leave a more detailed analysis for future work.

We have argued that for $r \ll 0$, for odd $k$,
there is no Higgs branch contribution
to the GLSMs for symplectic Grassmannians,
and for even $k$, there is a Higgs branch.
In the full quantum theory, 
there is a Coulomb branch contribution,
just as in GLSMs for ordinary Grassmannians, and we will see explicitly
that those Coulomb vacua provide an explicit representation of
the quantum cohomology ring.  Furthermore, the Higgs branch (if it exists)
will also supply the difference between the Witten index of the
$r \gg 0$ phase and the number of Coulomb vacua (corresponding to roots
of the ring relations).

In the next section we will describe how the Coulomb branch realizes
both ordinary and equivariant quantum cohomology of
general Lagrangian Grassmannians, and also check in special
cases that the quantum cohomology ring of other symplectic Grassmannians
is also realized by these GLSMs.

\subsection{Quantum cohomology of Lagrangian Grassmannians}

\subsubsection{Ordinary quantum cohomology}

In this section, we will argue that the physical chiral ring of this
theory coincides with known results for the quantum cohomology of
$SG(n,2n)$, which serves as a consistency check on the GLSM description
above.

We begin by studying the Coulomb branch.
The effective twisted superpotential is 
\begin{align*}
	\widetilde{W}_{eff} =& -t\sum_{a=1}^n \Sigma_a - \sum_{i=1}^{2n} \sum_{a,b=1}^n \rho_{ia}^b \Sigma_b\left[\ln \left(\sum_{b=1}^n \rho_{ia}^c \Sigma_c\right) -1\right] -
	 \sum_{\mu , \nu=1}^n \sum_{a=1}^n \rho^a_{\mu\nu} \Sigma_a \left[\ln\left( \sum_{b=1}^n\rho^b_{\mu\nu} \Sigma_b \right) - 1 \right] \\
	 &\quad - \sum_{\mu,\nu = 1}^n\sum_{a=1}^n\alpha_{\mu\nu}^a \Sigma_a \left[ \ln\left(\sum_{b=1}^n\alpha_{\mu\nu}^b\Sigma_b\right) - 1 \right] ,
\end{align*}
where $\rho_{ib}^a = \delta_b^a$, $\rho^a_{\mu\nu} = -\delta^a_\mu - \delta^a_\nu$ and $\alpha_{\mu\nu}^a = -\delta_\mu^a + \delta_\nu^a$. 
Here, the weights $\rho^a_{\mu\nu}$ correspond to $q_{ab}$ in the 
representation $\wedge^2V^*$ and can be obtained from the weights for $V^*$ following appendix~\ref{app:product}. 
One can check that the $\rho^a_{\mu\nu}$ make the 
superpotential~(\ref{eq:lagsuperpotential}) gauge invariant. 
Note that the sum of the W-boson contributions contributes a $i(n-1)\pi$-shift 
to $t$.  Computing the critical locus, we have
\begin{equation}
	\frac{\partial \widetilde{W}_{eff}}{\partial \sigma_a } = - t - i (n-1) \pi  - 2n \ln \sigma_a + \sum_{b \neq a} \ln(-\sigma_a - \sigma_b)
\end{equation}
which gives the chiral ring relations 
\begin{equation}
\label{eq:sggeneral}
	q \prod_{b \neq a}(\sigma_a+\sigma_b) = \sigma_a^{2n}, \ {\rm for}\ a = 1, \cdots,n,
\end{equation}
with (from the ambient theory) excluded locus $\sigma_a \neq \sigma_b$ if $a \neq b$. 

Mathematically,
the quantum cohomology ring relations for $SG(n,2n)$ are known, and can
be found in e.g. \cite[equ'n (3)]{kt2}:
\begin{equation}
\label{eq:sgqchrel}
	e_i(x)^2 + 2 \sum_{k=1}^{n-i} (-1)^k e_{i+k}(x)e_{i-k}(x) = (-)^{i+1}e_{2i-n-1}(x) \tilde{q}, \quad {\rm for}\quad i = 1,2,\dots,n.
\end{equation}
(To avoid symbol abuse, we have used $e_i$'s and $\tilde{q}$ here, 
instead of $\sigma_i$'s and $q$ as in \cite{kt2}.)
In equation~(\ref{eq:sgqchrel}), 
$e_i(x)$ is the $i$-th elementary symmetric polynomial in the Chern roots 
$\{ x_1, \dots, x_n\}$ of the tautological\footnote{
Reference \cite{kt2} works with Chern roots of ${\cal Q}$. 
In this specific example, the choice of either ${\cal Q}$ or ${\cal S}$ 
is equivariant as ${\cal S}^{*} = {\cal Q}$. 
However, different choices will have different $(-1)$ factors in 
the right-hand-side of (\ref{eq:sgqchrel}).} 
bundle ${\cal S}$, which can be understood as the restriction
of the tautological bundle on the ambient $G(n,2n)$ to 
$SG(n,2n)$, and so fits in the short exact sequence
\begin{equation}
	0 \longrightarrow {\cal S} \longrightarrow {\cal V}_{SG(n,2n)} 
\longrightarrow 
{\cal Q} \longrightarrow 0.
\end{equation}

We will argue in this section 
that the physical chiral ring relations~(\ref{eq:sggeneral}) 
reproduce the quantum cohomology ring relations~(\ref{eq:sgqchrel}) known
in the mathematics literature for
\begin{equation}
\label{eq:id}
\bigg\{
\begin{array}{l}
	q = (-)^{n-1}\tilde{q},\\
	\sigma_a = - x_a.
\end{array}
\end{equation}
These two identifications make sense in the following ways. 
First, as in \cite{Witten:1993xi,Guo:2015caf,Guo:2018iyr}, 
we can interpret the $\sigma_a$ as the Chern roots of ${\cal S}^*$, 
hence $\sigma_a = - x_a$. 
From naively counting degrees, one finds that both $q$ and $\tilde{q}$
have degree $n+1$, and they should match up to a constant factor,
which one can show is
$(-)^{n-1}$. Before giving a general argument that the physical chiral
ring relations match the ring relations known in mathematics,
we demonstrate how this works in several examples.

\noindent \underline{$n=1$}

As mentioned earlier, $SG(1,2)\simeq \mathbb{CP}^1$. 
The physical chiral ring relation, equation~(\ref{eq:sggeneral}), 
is $\sigma^2 = q$, which implies the quantum cohomology ring 
relation from equation~(\ref{eq:sgqchrel}),
\begin{equation}
	e_1(x)^2 = \tilde{q}.
\end{equation}
if equation~(\ref{eq:id}) is satisfied. Here, $e_1(x) = x$.

\noindent \underline{$n=2$}

$SG(2,4)$ is the first nontrivial example. In this example, 
equation~(\ref{eq:sggeneral}) gives
\begin{equation}
	q (\sigma_1+\sigma_2) = \sigma_1^4, \quad q (\sigma_1+\sigma_2) = \sigma_2^4.
\end{equation}
The mathematical quantum cohomology ring relations in this case are
\begin{equation}
	e_1(x)^2 = 2 e_2(x), \quad e_2(x)^2 = - \tilde{q} e_1(x).
\end{equation}

To see how the mathematical relations follow from the physical relations,
we
subtract the two physical chiral ring relations to get
\begin{displaymath}
\left(\sigma_1^2 - \sigma_2^2 \right)\left(\sigma_1^2 + \sigma_2^2 \right)
= 0. 
\end{displaymath}
Taking into account the excluded locus
$\sigma_1 \neq \sigma_2$, 
$\left(\sigma_1^2 - \sigma_2^2 \right)$ can be factored out and the above equation becomes
\begin{equation}
	\sigma_1^2 + \sigma_2^2 = 0,
\end{equation}
which is the same as $e_1(x)^2 = 2 e_2(x)$ with $\sigma_a = - x_a$.

Similarly, the sum of the two physical chiral ring relations is
\begin{equation}
2q\left(\sigma_1 + \sigma_2 \right) = \sigma_1^4 +\sigma_2^4 = - 2 \sigma_1^2 \sigma_2^2,
\end{equation}
where the last equality follows from
\begin{equation}
0 \equiv (\sigma_1^2 + \sigma_2^2)^2 = \sigma_1^4 + \sigma_2^4 + 2 \sigma_1^2 \sigma_2^2.
\end{equation}
Therefore, we have 
\begin{equation}
\sigma_1^2 \sigma_2^2 = - q \left(\sigma_1 + \sigma_2 \right).
\end{equation}
This is the same as $e_2(x)^2 = - \tilde{q} e_1(x)$ with $\sigma_a = - x_a$ and $\tilde{q} = -q$.

\noindent \underline{General case}

In the cases above, we used algebraic tricks 
to construct Weyl invariant polynomials of the $\sigma_a$, 
which led to the quantum cohomology ring relations. 
We will next use more general methods to study the cases $n \geq 3$.

First, note that the left-hand side of equation~(\ref{eq:sggeneral}) 
can be expanded in terms of Weyl invariant polynomials of 
the $\sigma_a$. For example, when $n=2k+1$, it can be expanded as:
\begin{equation}
\label{eq:noddchiral}
	q \sigma_a^{2k} + q e_2(\sigma) \sigma_a^{2k-2} + \dots + q e_{2k-2}(\sigma) \sigma_a^2 + q e_{2k}(\sigma) \: = \: \sigma_a^{4k+2},
\end{equation}
where $e_i(\sigma)$ is the $i$-th elementary symmetric polynomial.
Similarly, for $n=2k$, it can be rewritten as
\begin{equation}
\label{eq:noevenchiral}
	q e_1(\sigma) \sigma_a^{2k-2} + q e_3(\sigma) \sigma_a^{2k-4} + \dots + q e_{2k-3}(\sigma) \sigma_2 + q e_{2k-1}(\sigma) \: = \: \sigma_a^{4k}.
\end{equation}

Let us consider the $n=2k+1$ case first. 
Since the $e_i(\sigma)$ are Weyl invariant, 
the $e_i(\sigma)$ are constant on Weyl orbits.
Rewrite equation~(\ref{eq:noddchiral}) as
\begin{equation} \label{eq:sgn2n:roots}
	P(\sigma_a^2) \: \equiv \: (\sigma_a^2)^{2k+1} - q (\sigma_a^2)^{k} - q e_2(\sigma) (\sigma_a^2)^{k-1} - \dots - q e_{2k-2}(\sigma) \sigma_a^2 - q e_{2k}(\sigma) \: = \: 0, 
\end{equation}
for $a=1, \dots, n$. Then the $\sigma_a^2$ satisfy relations determined by 
the coefficients of $P(\sigma_a^2)$ according to Vieta's formula
\cite[theorem 2]{blum-coskey},
which says that any $n$th order polynomial $p(z)$ with roots $a_1, \cdots, a_n$
can be written in the form
\begin{equation}
p(z) \: \propto \:
z^n - e_1(a) z^{n-1} + e_2(a) z^{n-2} + \cdots + (-)^n e_n(a),
\end{equation}
where the $e_i(a)$ are elementary symmetric polynomials in the roots $a_j$.

If we let $x_a$ denote a root of equation~(\ref{eq:sgn2n:roots}),
a solution for $\sigma_a^2$,
then from Vieta's formula and the coefficients of the $\sigma^2$ terms
in~(\ref{eq:sgn2n:roots}), we have   
\begin{equation}
\begin{array}{*1{>{\displaystyle}c}p{6cm}}
	\sum_{1\leq a \leq n} x_a = 0,\\
	\sum_{1\leq a_1<a_2 \leq n}x_{a_1}x_{a_2} = 0,\\
	\cdots \\
	\sum_{1\leq a_1<\dots < a_k \leq n}x_{a_1}\dots x_{a_k} =  0,
\end{array}
\end{equation}
and 
\begin{equation}
\begin{array}{*1{>{\displaystyle}c}p{6cm}}
	\sum_{1\leq a_1<\dots < a_{k+1} \leq n}x_{a_1}\dots x_{a_{k+1}} = (-1)^{k} q,\\
	\sum_{1\leq a_1<\dots < a_{k+2} \leq n}x_{a_1}\dots x_{a_{k+2}} = (-1)^{k+1} q e_2(\sigma),\\
	\cdots \\
	\sum_{1 \leq a \leq n}x_{1}\dots\widehat{x_a} \dots x_{2k+1} = (-1)^{2k+1} q e_{2k-2}(\sigma) ,\\
	x_{1}\dots x_{2k+1} = (-1)^{2k+2} q e_{2k}(\sigma).
\end{array}
\end{equation}
The reader should note that the number of possible roots ($2k+1$, the degree
of equation~(\ref{eq:sgn2n:roots}), matches the rank ($n$) of the gauge
group.
The equations above (derived from the coefficients of
equation~(\ref{eq:sgn2n:roots}) can be written more compactly
for $\ell \leq n = 2k+1$
as
\begin{equation}  \label{eq:sgn2n:compact1}
\sum_{1 \leq a_1 < \cdots < a_{\ell} \leq n} \sigma_{a_1}^2 \cdots
\sigma_{a_{\ell}}^2 \: = \: (-)^{\ell-1} e_{2\ell - n - 1}(\sigma) q,
\end{equation}
in conventions in which $e_i = 0$ for $i < 0$ and $e_0 = 1$.
Applying identity~(\ref{eq:app:symm:e1}), which we repeat below,
\begin{equation}
\label{eq:identityee}
        e_{\ell}(\sigma)^2 + 2 \sum_{j=1}^{n-l} (-1)^j e_{\ell+j}(\sigma) e_{\ell-j}(\sigma) 
= \sum_{1 \leq a_1 < \dots < a_{\ell} \leq n}\sigma_{a_1}^2\dots\sigma_{a_{\ell}}^2,
\end{equation}
we recover the quantum cohomology ring relations for $SG(2k+1,4k+2)$
known in the math community,
equation~(\ref{eq:sgqchrel}),
with $\sigma_a = - x_a$ and $q = (-)^{2k}\tilde{q} = \tilde{q}$.

The same argument works for $n=2k$. 
Rewrite equation~(\ref{eq:noevenchiral}) as 
\begin{equation}
	P(\sigma_a^2) \equiv (\sigma_a^2)^{2k} - q e_1(\sigma) (\sigma_a^2)^{k-1} - \dots - q e_{2k-3}(\sigma) \sigma_a^2 - q e_{2k-1}(\sigma) \: = \: 0,
\end{equation}
then Vieta's formula and the coefficients of the polynomial
above give the relations
\begin{equation}
\begin{array}{*1{>{\displaystyle}c}p{6cm}}
	\sum_{1\leq a \leq n} \sigma_a^2 = 0,\\
	\sum_{1\leq a_1<a_2 \leq n}\sigma_{a_1}^2\sigma_{a_2}^2 = 0,\\
	\dots \\
	\sum_{1\leq a_1<\dots < a_k \leq n}\sigma_{a_1}^2\dots\sigma_{a_k}^2 = 0,
\end{array}
\end{equation}
and
\begin{equation}
\begin{array}{*1{>{\displaystyle}c}p{6cm}}
	\sum_{1\leq a_1<\dots < a_{k+1} \leq n}\sigma_{a_1}^2\dots\sigma_{a_{k+1}}^2 = (-1)^{k} q e_1(\sigma),\\
	\sum_{1\leq a_1<\dots < a_{k+2} \leq n}\sigma_{a_1}^2\dots\sigma_{a_{k+2}}^2 = (-1)^{k+1} q e_3(\sigma),\\
	\dots \\
	\sum_{1 \leq a \leq n}\sigma_{1}^2\dots\widehat{\sigma_a^2} \dots \sigma_{2k}^2 = (-1)^{2k} q e_{2k-3}(\sigma),\\
	\sigma_{1}^2\dots\sigma_{2k}^2 = (-1)^{2k+1} q e_{2k-1}(\sigma),
\end{array}
\end{equation}
where here we have abused notation by labeling the possible solutions for 
$\sigma^2$ by
$\sigma^2$ instead of $x$.  Just as before, these can be summarized
compactly as in equation~(\ref{eq:sgn2n:compact1}),
and applying identity~(\ref{eq:app:symm:e1}),
we recover
the quantum cohomology ring relations for $SG(2k,4k)$, 
equation~(\ref{eq:sgqchrel}).

So far we have checked our previous claim that the chiral ring  
relations~(\ref{eq:sggeneral}) in the Coulomb branch
do reproduce the quantum cohomology ring relations~(\ref{eq:sgqchrel}).

\subsubsection{Equivariant quantum cohomology}

The above story can be generalized to the equivariant case, 
equivariant with respect to a maximal torus of the flavor symmetry group,
which physically corresponds to turning on twisted masses \cite{Hori:2000kt}. 
We will verify that the equivariant quantum cohomology ring predicted by
the physics of this GLSM matches that known in mathematics for Lagrangian
Grassmannians.

Due to the global symmetry $Sp(2n)$, 
which preserves the isotropy  
 condition~(\ref{eq:sgisotropy}), 
we have $m_{-i} = - m_i$. 
To keep the superpotential invariant, the twisted masses for $q_{ab}$ is 
\begin{equation}
	m_q \: = \: - \sum_{i=1}^n (m_i + m_{-i})
\: = \: - \sum_{i=1}^n (m_i - m_{i}) \: = \: 0.
\end{equation}
Therefore, the effective twisted superpotential becomes
\begin{align*}
	\widetilde{W}_{eff} =& -t\sum_{a=1}^n \Sigma_a - \sum_{a=1}^n \sum_{i=\pm 1}^{\pm n} \left(\Sigma_a-m_i\right)\left[\left(\ln \Sigma_a -m_i\right) -1\right] \\
		&\quad + \sum_{\mu > \nu=1}^n \sum_{a=1}^n \rho^a_{\mu\nu} \Sigma_a \left[\ln\left( - \sum_{b=1}^n\rho^b_{\mu\nu} \Sigma_b \right) - 1 \right] \\
		&\quad - \sum_{\mu,\nu = 1}^n\sum_{a=1}^n\alpha_{\mu\nu}^a \Sigma_a \left[ \ln\left(\sum_{b=1}^n\alpha_{\mu\nu}^b\Sigma_b\right) - 1 \right] ,
\end{align*}
and the critical locus of this effective twisted superpotential gives 
the following physical chiral ring relations
\begin{equation}
\label{eq:sgcringequiv}
	q \prod_{b \neq a}(\sigma_a+\sigma_b)
\: = \: \prod_{i=1}^n\left(\sigma_a^2 - m_i^2\right), \ {\rm for}\ a = 1, \cdots, n.
\end{equation}
We will show that these reproduce the equivariant quantum cohomology ring 
relations~(\ref{eq:sgqringeqiv}) in appendix~\ref{app:sg}. 
We repeat them here for convenience:
\begin{equation}
	e_{i}^2(x) + 2 \sum_{l=1}^{n-i} (-)^{l} e_{i-l}(x) e_{i+l}(x)
\: = \: e_{i}(t^2)+ (-)^{i+1}  e_{2i-n-1}(x) \tilde{q}. \nonumber
\end{equation}

Note that the right-hand-side of equation~(\ref{eq:sgcringequiv}) can be expanded as
\begin{equation}
	\prod_{i=1}^n\left(\sigma_a^2 - m_i^2\right)
\: = \: \sum_{i=0}^n (-1)^i \left(\sigma_a^2\right)^{n-i} e_i(m^2),
\end{equation}
while the left-hand-side of equation~(\ref{eq:sgcringequiv}) 
can be expanded in the same way as before. 
To establish that the mathematical ring is a consequence of the
physical chiral ring, we can use the same methods as in our analysis of
the ordinary quantum cohomology.
Here for brevity we will only give the details for the $n=2k+1$ case.
First, we rewrite equation~(\ref{eq:sgcringequiv}) as
\begin{eqnarray}
\lefteqn{
q \left(\sigma_a^2\right)^{k} + q e_2 (\sigma) \left(\sigma_a^2\right)^{k-1} + \dots + q e_{2k-2}(\sigma) \left(\sigma_a^2\right) + q e_{2k}(\sigma)
} \nonumber \\
& \hspace*{1.0in} = & \sum_{i=0}^{2k+1} (-1)^i \left(\sigma_a^2\right)^{2k+1-i} e_i(m^2),
\end{eqnarray}
for $a = 1,\dots, n$, 
where $e_i(m^2)$ is the $i$-th elementary symmetric polynomial of $\{ m_1^2, \cdots, m_n^2 \}$. 
Following the same reasoning as before, we choose to work on one vacuum, 
corresponding to one Weyl orbit of solutions for the $\sigma_a$, 
and on this orbit the $e_{i}(\sigma)$ are constant due to Weyl invariance. 
Therefore, the equation above is of degree $(2k+1)$ in $\sigma_a^2$.

From Vieta's formula, we have following sets of relations:
\begin{align*}
	\sum_{1\leq a \leq n} \sigma_a^2 = e_1(m^2),\\
	\sum_{1\leq a_1<a_2 \leq n}\sigma_{a_1}^2\sigma_{a_2}^2 = e_2(m^2),\\
	\dots \\
	\sum_{1\leq a_1<\dots < a_k \leq n}\sigma_{a_1}^2\dots\sigma_{a_k}^2 = e_k(m^2),
\end{align*}
and
\begin{align*}
	\sum_{1\leq a_1<\dots < a_{k+1} \leq n}\sigma_{a_1}^2\dots\sigma_{a_{k+1}}^2 &= e_{k+1}(m^2)+(-1)^{k} q,\\
	\sum_{1\leq a_1<\dots < a_{k+2} \leq n}\sigma_{a_1}^2\dots\sigma_{a_{k+2}}^2 &= e_{k+2}(m^2)+(-1)^{k+1} q e_2(\sigma),\\
	\dots & \\
	\sum_{1 \leq a \leq n}\sigma_{1}^2\dots\widehat{\sigma_a^2} \dots \sigma_{2k+1}^2 &=  e_{2k}(m^2) - q e_{2k-2}(\sigma),\\
	\sigma_{1}^2\dots\sigma_{2k+1}^2 &= e_{2k+1}(m^2) + q e_{2k}(\sigma).
\end{align*}
From equation~(\ref{eq:identityee}), we can summarize the equations above as
\begin{equation}
	e_{i}^2(\sigma) + 2 \sum_{l=1}^{2k+1-i} (-)^{l} e_{i-l}(\sigma) e_{i+l}(\sigma) 
\: = \: e_{i}(m^2)+ (-)^{i+1} q e_{2i-2k-2}(\sigma),
\end{equation}
which match the known mathematical results for equivariant
quantum cohomology, equation~(\ref{eq:sgqringeqiv}), for  
\begin{equation}
\left\{
\begin{array}{l}
	q \: = \:  (-)^{n-1}\tilde{q},\\
	\sigma_a \: = \: - x_a,\\
	m_i^2 \: =  \: t_i^2.
\end{array}\right.
\end{equation}

\subsection{Quantum cohomology for general symplectic Grassmannians}
\label{sec:gsp}

In this section, we will compare quantum cohomology of general
symplectic Grassmannians $SG(k,2n)$ to the predictions of the physical
chiral ring.  We will not give a general proof that they always match,
but instead will merely check several families of examples.

The analysis in the previous section can be applied to $SG(k,2n)$.
The physical chiral ring relations have the same form as before, 
but now there are only $k < n$ relations:
\begin{equation}
\label{eq:sgkncring}
	q\prod_{b \neq a}(\sigma_a + \sigma_b) 
\: = \: \sigma_a^{2n}, \quad {\rm for }\quad a = 1, \dots, k.
\end{equation}

These chiral ring relations will reproduce the quantum cohomology ring relations \cite{buch2009quantum}, which are
\begin{equation}
\label{eq:sgknqring}
	c_{r}^{2}+2 \sum_{i=1}^{2n-k-r}(-1)^{i} c_{r+i} c_{r-i}
\: = \:
(-1)^{2n-k-r} c_{2 r +k -2 n-1} \tilde{q}, \quad n-k+1 \leq r \leq n,
\end{equation}
where $c_r$ is the $r$-th Chern class of the quotient bundle ${\cal Q}$ 
over $SG(k,2n)$, defined by restricting the universal quotient bundle
over the ambient $G(k,2n)$. 
It obeys
\begin{equation}
	0 \: \longrightarrow \: {\cal S} \: \longrightarrow \: {\cal V}_{SG(k,2n)} 
 \: \longrightarrow \: {\cal Q} \: \longrightarrow \: 0,
\end{equation}
hence we have $c({\cal Q})c({\cal S}) = 1$ which implies 
\begin{equation}
	c_r({\cal Q}) \: = \: (-)^r\det\left(c_{1+j-i}({\cal S})\right)_{1\leq i,j \leq r}.
\end{equation}
If we interpret the $\sigma_a$ as the Chern roots of ${\cal S}^{*}$, 
then
we have $c_{i}({\cal S}) = (-)^i e_{i}(\sigma)$, and 
\begin{equation}
	c_r({\cal Q}) \: =  \: h_r(\sigma),
\end{equation}
where $h_r(\sigma)$ is the $r$th complete homogeneous symmetric polynomial 
in the $\sigma_a$.

We will check in a series of examples that physics correctly
reproduces the quantum cohomology ring relations above.

First, recall that for $k=1$, the isotropy condition is trivially satisfied,
hence $SG(1,2n) \simeq G(1,2n)\simeq {\mathbb P}^{2n-1}$. 
If we set $k=1$ in equation~(\ref{eq:sgkncring}), we get
\begin{equation}
	\sigma^{2n} \: = \: q,
\end{equation}
which is indeed the chiral ring relation for ${\mathbb P}^{2n-1}$. 
At the same time, equation~(\ref{eq:sgknqring}) for $k=1$ is just
\begin{equation}
	c_n^2 \: = \: (-)^{n-1} \tilde{q},
\end{equation}
where $c_{n} = h_n (\sigma) = \sigma^n$. 
Therefore, the ring relations match if we identify
\begin{displaymath}
q \: = \: (-)^{n-1}\tilde{q}.
\end{displaymath}
This identification of $q$ and $\tilde{q}$ is generally true as we will see in the following examples.

Next we consider the case $SG(2,6)$.
The physical chiral ring relations in this case are
\begin{align*}
	&q (\sigma_1+\sigma_2) \: = \: \sigma_1^6,\\
	&q (\sigma_1+\sigma_2) \: = \: \sigma_2^6.
\end{align*}
Some algebraic manipulations give
\begin{align*}
	 & \sigma_1^4 + \sigma_1^2 \sigma_2^2 + \sigma_2^4 \: = \: 0, \\
	 & \sigma_1^6 + \sigma_2^6 \: = \: 2 q (\sigma_1+\sigma_2),
\end{align*}
and these two Weyl invariant equations indeed reproduce 
equation~(\ref{eq:sgknqring}) for quantum cohomology when $k=2$:
\begin{equation}
	c_r^2+2\sum_{i=1}^{4-r}(-1)^i c_{r+i}c_{r-i} \: =  \: (-1)^{4-r}c_{2r-5}\tilde{q}, \quad \quad 2\leq r\leq 3,
\end{equation}
with $q = (-)^{3-1}\tilde{q} = \tilde{q}$. 
Note that we have used equation~(\ref{eq:cspidentity}) in appendix~\ref{app:symm}.

Our next example is $SG(2,8)$. The physical chiral ring relations are
\begin{align*}
	&q (\sigma_1+\sigma_2)  \: = \: \sigma_1^8,\\
	&q (\sigma_1+\sigma_2) \: = \: \sigma_2^8,
\end{align*}
which yield
\begin{align*}
	 & \sigma_1^6+\sigma_1^4 \sigma_2^2+\sigma_1^2 \sigma_2^4+\sigma_2^6 = 0, \\
	 & \sigma_1^8 + \sigma_2^8 \: = \: 2 q (\sigma_1+\sigma_2).
\end{align*}
We can check that these two equations reproduce equation~(\ref{eq:sgknqring}) 
when $k=2$ if $q = (-)^{4-1}\tilde{q}$. 

This calculation can be generalized to $SG(2,2n)$ and,
due to the equation~(\ref{eq:cspidentity}), 
the chiral ring relations for $SG(2,2n)$ 
reproduce the quantum cohomology ring relations.

\subsection{Witten indices}
\label{sect:sg:witten}

As a consistency check of our description and analysis, here we will check
that Witten indices are preserved across various phases of these
GLSMs.  (Although the target is not Calabi-Yau, and so the axial R-symmetry
is anomalous, nevertheless, there can be nonanomalous finite subgroups,
and so one expects continuous paths connecting the $r \gg 0$ and
$r \ll 0$ phases of these GLSMs, hence the Witten indices should match.)

For simplicity, let us begin
with the case $LG(2,4)$.  As previously described, the Euler
characteristic of this space is $2^2 = 4$, which should match the
Witten index of the $r \gg 0$ phase, if our GLSM is correct.
This should also be the Witten index of the $r \ll 0$ phase, which we
will now check.

The $r \ll 0$ phase is a mixed Higgs/Coulomb branch.  As previously
discussed in section~\ref{sect:sg:background}, the Higgs branch at low energies
is an $SU(2)$ gauge theory with $4$ fundamentals, which has Witten index
$1$.  The Coulomb branch is defined by solutions to the equations
to the quantum cohomology ring relations
\begin{equation}
e_1(x)^2 - 2 e_2(x) e_0(x) \: = \: 0, \: \: \:
e_2(x)^2 \: = \: - e_{1}(x) \tilde{q},
\end{equation}
or after simplification,
\begin{equation}
x_1^2 + x_2^2 \: = \: 0, \: \: \:
x_1^2 x_2^2 \: = \: - (x_1 + x_2) \tilde{q}.
\end{equation}
It is straightforward to check that there are $3$ distinct unordered
pairs $(x_1, x_2)$ which solve these equations (and are consistent with
the excluded locus), hence there are $3$ Coulomb vacua.
The sum of the number of Coulomb vacua and the Witten index of the Higgs
branch is $4$, matching the Witten index of the $r \gg 0$ phase,
as expected.

Next, consider $SG(2,2n)$.  The Euler characteristic of this space,
the Witten index of the phase $r \gg 0$, is 
\begin{equation}
2^2 \binom{n}{2} 
 \: = \: 2 n (n-1).
\end{equation}
From the general analysis of
section~\ref{sect:sg:background}, the Higgs branch of the 
$r \ll 0$ phase has Witten index $n-1$, so to be consistent, there
should be
\begin{equation}
2n (n-1) - (n-1) \: = \: (2n-1)(n-1)
\end{equation}
Coulomb vacua.

We count the Coulomb vacua of the $r \ll 0$ phase of the GLSM for
$SG(2,2n)$ as follows.  The chiral ring relations are
\begin{equation}
\sigma_1^{2n} \: = \: q (\sigma_1 + \sigma_2) \: = \: 
\sigma_2^{2n},
\end{equation}
and the (unordered) roots of these equations are the Coulomb vacua
(subject to the excluded locus $\sigma_1 \neq \pm \sigma_2$,
which requires $\sigma_a \neq 0$).  Since $\sigma_1^{2n} = \sigma_2^{2n}$,
we see that $\sigma_1$ and $\sigma_2$ differ by a $2n$-th root of unity
(excluding $-1$, as that is on the excluded locus). 
Excluding $\pm 1$, we see that $\sigma_1$ differs from $\sigma_2$ by 
$2n-2$ possible phases.  Plugging in, the chiral ring relations reduce
to a degree $2n-1$ polynomial in either of the $\sigma_a$, hence
$2n-1$ solutions.  Dividing by $2$ to account for ordering, we have a total
of
\begin{equation}
\frac{1}{2} (2n-2)(2n-1) \: = \: (2n-1)(n-1)
\end{equation}
possible Coulomb vacua, which is precisely right for the Witten index
of the $r \gg 0$, $r \ll 0$ phases of $SG(2,2n)$ to match.

To the same end, let us now discuss $SG(k,2n)$ with $k$ odd.
Here, as previously argued, there is no Higgs branch, only a Coulomb branch,
hence for consistency the number of Coulomb vacua should match the
Euler characteristic of the space, which is
\begin{equation}
2^k \binom{n}{k}. 
\end{equation}
We can see this as follows.
The chiral ring relations with $k$ odd are
\begin{equation}
\label{eq: chiralring01}
        q \prod_{b\neq a}(\sigma_a + \sigma_b ) \: = \: \sigma_a^{2n},
\end{equation}
for all $a=1,\dots,k$. Equation~(\ref{eq: chiralring01}) can be 
rewritten in the form
\begin{equation}
\label{eq:chiralring02}
        \sigma_a^{2n} - q \left[ \sigma_a^{k-1} + e_2(\sigma) \sigma_a^{k-3} + \cdots + e_{k-1}(\sigma) \right] \: = \: 0.
\end{equation}
Since $k$ is odd, we know that if $\{\sigma_1,\dots,\sigma_k\}$ is one 
solution, then $\{-\sigma_1,\dots,-\sigma_k\}$ is another solution. 
As we are counting vacua away from the excluded locus
($\sigma_a \neq 0$, $\sigma_a \neq \pm \sigma_b$ for $a \neq b$) 
the solutions should always have this $\mathbb{Z}_2$ symmetry.
Note that this is not true for cases in which $k$ is even, because  
elementary symmetric polynomials of odd degrees appear.

Putting this together,
there are $2n$ choices for $\sigma_1$, which leaves $2n-2$ choices for 
$\sigma_2$, as $\sigma_2 \neq \pm \sigma_1$.  Continuing,
one eventually finds $2n-2k+2$ choices for $\sigma_k$. 
Therefore, there are in total $2n(2n-2)\cdots(2n-2k+2)$ choices. 
We also divide by $k!$ to remove permutations, which gives
\begin{equation}
 \frac{2n(2n-2)\cdots(2n-2k+2)}{k!} \: = \: 2^k \binom{n}{k}
\end{equation}
Coulomb vacua, as expected in order for Witten indices to match.

In $SG(k,2n)$ for $k$ odd, as previously discussed, there is no Higgs
branch at $r \ll 0$, only Coulomb vacua.  For $k$ even, there can be
a nontrivial Higgs branch, which contributes to the Witten index,
as discussed in greater detail previously.

In table~\ref{table:witten-check} we have summarized results for a number
of cases, comparing Euler characteristics of large-radius phases,
number of Coulomb vacua, and Witten indices of Higgs branches at
$r \ll 0$.  In each case, we find that the large-radius Euler characteristic
matches the sum of the number of Coulomb vacua and the Witten index
of the Higgs branch.

\begin{table}[ht]
\centering
\begin{tabular}{c|ccc}
Geometry & $\chi$(geometry) & Num. Coulomb vacua & $\chi$(Higgs) \\ \hline
$SG(1,2n)$ & $2n$ & $2n$ & $0$ \\
$SG(2,4)$ & $4$ & $3$ & $1$ \\
$SG(2,6)$ & $12$ & $10$ & $2$ \\
$SG(2,8)$ & $24$ & $21$ & $3$ \\
$SG(3,6)$ & $8$ & $8$ & $0$ \\
$SG(3,8)$ & $32$ & $32$ & $0$ \\
$SG(3,10)$ & $80$ & $80$ & $0$ \\
$SG(4,8)$ & $16$ & $15$ & $1^*$ \\
$SG(4,10)$ & $80$ & $77$ & $3^*$
\end{tabular}
\caption{Listed are Euler characteristics of large radius phases,
number of Coulomb vacua, and Witten indices of Higgs branches of
$r \ll 0$ phases.  In each case, the sum of the number of Coulomb
vacua and Higgs Witten indices matches the large-radius Euler
characteristic, as expected.  Euler characteristics of geometries
are computed using the exact expression in section~\ref{sect:sg:background}, and
Coulomb vacua were counted either analytically or, in some
cases, numerically.  Euler characteristics of Higgs
branches are as given in section~\ref{sect:sg:background}.  
For the latter, we only have an
exact result for cases $k$ odd and $k=2$.  For the case $k=4$, we made
a conjecture in section~\ref{sect:sg:background}, 
whose result we list here.  We denote
conjectured results with an asterisk ($*$), and observe that they happen
to have the correct values to preserve Witten indices.
\label{table:witten-check}}
\end{table}

\subsection{Calabi-Yau condition}

As another consistency test, we briefly mention Calabi-Yau conditions.
Mathematically, the intersection of the Pl\"ucker embedding of
$SG(k,2n)$ with a hypersurface of degree $2n-k+1$ hypersurface is
Calabi-Yau.  We reproduce the same condition physically as
the condition for the sum of the charges under any
$U(1)$ subgroup of the gauge group to vanish.

The GLSM for $SG(k,2n)$ is a $U(k)$ gauge theory with $2n$ chirals
in the fundamental $V$, and one chiral in $\wedge^2 V^*$.
Under any $U(1) \subset U(k)$, the $2n$ chirals in the fundamental
contribute a total of $2n$ to the sum of the $U(1)$ charges,
and the one chiral in $\wedge^2 V^*$ contributes $-(k-1)$,
so that the sum of the $U(1)$ charges is
\begin{equation}
2n - k + 1.
\end{equation}
Under the same $U(1)$, any element of the Pl\"ucker embedding
\begin{equation}
\epsilon_{a_1 \cdots a_k} \phi^{a_1}_{i_1} \cdots
\phi^{a_k}_{i_k}
\end{equation}
has charge $1$, so we see that intersecting the image of $SG(k,2n)$
with a hypersurface of degree $2n-k+1$ should be Calabi-Yau, reproducing
the mathematics
result. 

Let us consider two special cases as explicit confirmations.
\begin{itemize}
\item $SG(1,2n) = {\mathbb P}^{2n-1}$.  The condition for a hypersurface
in ${\mathbb P}^{2n-1}$ to be Calabi-Yau is that it has degree $2n$,
which is reproduced by the condition above.
\item $SG(2,4) = {\mathbb P}^4[2]$.  Here, the Calabi-Yau condition is that
a hypersurface should have degree $3$, which is reproduced by the 
condition above.
\end{itemize}

\subsection{Symplectic flag manifolds}

In addition to symplectic Grassmannians, there also exist
symplectic flag manifolds.  At the level of group cosets,
these are of the form $Sp(2n,{\mathbb C}) / P$ for suitable parabolic subgroups
$P$.  We can describe them as submanifolds of ordinary
flag manifolds $F(k_1, \cdots, k_p, 2n)$ ($k_1 < k_2 < \cdots$)
satisfying an isotropy condition
on the maximal vector space.

Let us briefly describe GLSMs for these flag manifolds.
We begin with the GLSM for an ordinary flag manifold
$F(k_1, \cdots, k_p, 2n)$ \cite{Donagi:2007hi}.
This is a 
\begin{equation}
U(k_1) \times U(k_2) \times \cdots \times U(k_p)
\end{equation}
gauge theory with bifundamentals $({\bf k_1}, {\bf \overline{k_2}})$,
$({\bf k_2}, {\bf \overline{k_3}})$, and so forth to
$({\bf k_{p-1}}, {\bf \overline{k_{p}}})$, along with $2n$ chirals in
representation ${\bf k_p}$ of $U(k_p)$, following \cite{Donagi:2007hi}.
To build the GLSM for a symplectic flag manifold $SF(k_1, \cdots, k_p, 2n)$,
we add a chiral superfield $q_{ab}$ transforming in
the $\wedge^2 {\bf \overline{k_p}}$ representation of $U(k_p)$,
along with a superpotential
\begin{equation}
W \: = \: \sum_{i=1}^n q_{ab} \Phi^a_i \Phi^b_{-i}.
\end{equation}
We only impose an isotropy condition on the last, maximal flag:  as all
other vector spaces in the flag are subspaces of the maximal flag,
this suffices to guarantee that all subspaces satisfy the isotropy condition.

We will not compute quantum cohomology rings from the GLSM here,
but mathematical discussions of quantum cohomology rings for
symplectic flag manifolds can be found in
\cite{kim}.

\subsection{Mirrors of symplectic Grassmannians}

In this section we will briefly discuss mirrors to these nonabelian
GLSMs, following the nonabelian mirror ansatz discussed in
\cite{Gu:2018fpm}.  (It should be noted that other notions of 
mirrors exist, with different UV presentations but apparently equivalent
IR physics, see \cite{r08,mr,rw,pr,spacek,kalashnikov}.)

The mirror to the GLSM for $SG(k,2n)$ is
a Landau-Ginzburg model defined by \cite{Gu:2018fpm}
\begin{itemize}
	\item chiral superfields $Y_{ia}$, $i\in \{ \pm 1,\cdots, \pm n \}$ and $a\in\{1,\cdots,k \}$,
	\item chiral superfields $U_{\mu\nu} = \exp(-V_{\mu\nu})$, mirror to $q_{\mu\nu}$, $\mu,\nu \in\{1,\cdots,k \}$,
	\item chiral superfields $X_{\mu\nu} = \exp(-Z_{\mu\nu})$, mirror to W-bosons, $\mu,\nu \in\{1,\cdots,k \}$,
	\item $\sigma_a$.
\end{itemize}
with superpotential 
\begin{align*}
	W = &\sum_a \sigma_a \left( \sum_i Y_{ia} - \sum_{\mu > \nu} \rho^a_{\mu\nu} \ln U_{\mu\nu} - \sum_{\mu\neq\nu} \alpha^a_{\mu\nu} \ln X_{\mu\nu} - t \right) \\
	&\quad + \sum_{i,a}\exp(-Y_{ia}) + \sum_{\mu>\nu}U_{\mu\nu} + \sum_{\mu\neq \nu} X_{\mu\nu},
\end{align*}
where $\rho_{\mu\nu}^a = - \delta^a_\mu - \delta^a_{\nu}$. 
Here we are considering general symplectic Grassmannians $SG(k,2n)$ 
with $k \leq n$, 
which includes Lagrangian Grassmannians as a special case when $k=n$.

Let us check explicitly that the chiral ring relations match. 
First, integrate out the $\sigma_a$ to get the constraints
\begin{equation}
	\sum_i Y_{ia} - \sum_{\mu > \nu} \rho^a_{\mu\nu} \ln U_{\mu\nu} - \sum_{\mu\neq\nu} \alpha^a_{\mu\nu} \ln X_{\mu\nu} \: = \: t,
\end{equation}
which we solve by taking
\begin{equation}
	Y_{na} \: = \: t - \sum_{i<n} Y_{ia} + \sum_{\mu > \nu} \rho^a_{\mu\nu} \ln U_{\mu\nu} + \sum_{\mu\neq\nu} \alpha^a_{\mu\nu} \ln X_{\mu\nu}.
\end{equation}
We define
\begin{equation}
	\Pi_a \: \equiv \: \exp(-Y_{na}) 
\: = \: q \left(\prod_{i<n}\exp(Y_{ia})\right)\left(\prod_{\mu \neq a}U_{a\mu}\right)\left(\prod_{\mu\neq a} \frac{X_{a\mu}}{X_{\mu a}}\right),
\end{equation}
then the superpotential reduces to
\begin{equation}
	W = \sum_{i<n,a}\exp(-Y_{ia})+ \sum_a \Pi_a + \sum_{\mu>\nu}U_{\mu\nu} + \sum_{\mu\neq \nu} X_{\mu\nu}.
\end{equation}
On the critical locus, we have
\begin{equation}
	\exp(-Y_{ia}) = \Pi_a,\quad - U_{\mu\nu} = \Pi_\mu + \Pi_\nu,\quad X_{\mu\nu} = - \Pi_\mu + \Pi_\nu.
\end{equation}
Therefore, the chiral ring relations are
\begin{equation}
	\Pi_a^{2n} = q \prod_{\mu\neq a} (\Pi_a + \Pi_\mu), \quad a = 1,\cdots,k.
\end{equation}

Now we can see that the mirror reproduces the chiral ring relation of the
original theory via the operator mirror map
\begin{equation}
	\Pi_a \leftrightarrow \sigma_a.
\end{equation}

\section{Orthogonal Grassmannians $OG(k,n)$}
\label{sect:og}

\subsection{Background and GLSM realization}

Orthogonal Grassmannians, denoted $OG(k,n)$,
are submanifolds of an ordinary
Grassmannian $G(k,n)$, satisfying an isotropy condition 
with respect to a nondegenerate quadratic form.
(Specifically, an isotropic subspace $W$ of a vector space $V$ has
the property that for all vectors $x, y \in W$, $x \cdot y=0$ for the
dot product defined by the quadratic form; nondegeneracy simply
means that the orthogonal complement of the entire vector space $V$ is
just $0$.)
They can be realized by GLSMs for Grassmannians with a superpotential realizing the
isotropy condition.  The resulting GLSMs look very similar to those for
symplectic Grassmannians, except that one has a field coupling to
a symmetric-tensor-square representation rather than in an antisymmetric
tensor representation.

We have two slightly different GLSMs depending upon whether $n$ is even or odd.
First consider the case of $n$ odd.  Write $n = 2m+1$.
The GLSM is a $U(k)$ gauge theory with $n$ chirals $\phi^a_i$
in the fundamental representation $V$
($a \in \{1, \cdots, k\}$, $i \in \{ -m, -m+1, \cdots, 0,
\cdots, +m\}$), and one chiral $q_{ab}$ in the representation Sym$^2 V^*$,
with superpotential
\begin{equation}
W \: = \: q_{ab} \left( \phi^a_0 \phi^b_0 \: + \: \sum_{i=1}^m \phi^a_i \phi^b_{-i} 
\right).
\end{equation}
We interpret $\phi^a_{\alpha}$ as defining $k$ vectors in ${\mathbb C}^{2m+1}$,
and the $F$ terms imply the isotropy condition $x \cdot y = 0$ for
each of $k$ vectors in ${\mathbb C}^{2m+1}$, with a dot product
defined by the symmetric matrix
\begin{equation}
\left[ \begin{array}{ccc}
1 & 0 & 0 \\
0 & 0 & I_m \\
0 & I_m & 0 \end{array} \right],
\end{equation}
corresponding to the quadratic form
\begin{equation}
Q(\phi) \: = \:  \phi_0 \phi_0 \: + \: \sum_{i=1}^m \phi_i \phi_{-i}.
\end{equation}

The case of $n$ even is similar.  Write $n=2m$.
The GLSM is a $U(k)$ gauge theory with $n$ chirals $\phi^a_i$ in the
fundamental representation $V$ ($a \in \{1, \cdots, k\}$, $i \in \{\pm 1, \pm 2, \cdots,
\pm m\}$), and one chiral $q_{ab}$ in the representation Sym$^2 V^*$,
with superpotential
\begin{equation}
W \: = \: \sum_{i=1}^m q_{ab} \phi^a_i \phi^b_{-i}.
\end{equation}
We interpret $\phi^a_{\alpha}$ as the components of $k$ vectors
in ${\mathbb C}^{2m}$, and the $F$ terms imply the isotropy condition
$x \cdot y = 0$ for each of $k$ vectors in ${\mathbb C}^{2m}$, with a dot
product defined by the symmetric matrix
\begin{equation}
\left[ \begin{array}{cc}
0 & I_m \\
I_m & 0 \end{array} \right],
\end{equation}
corresponding to the quadratic form
\begin{equation}
Q(\phi) \: = \:  \sum_{i=1}^m \phi_i \phi_{-i}.
\end{equation}

These GLSMs were also briefly described in
\cite[section 2.4.3, example 2]{ot}.

The dimension of $OG(k,n)$ (or rather, the dimension of one component,
in the case $n = 2k$) is\footnote{
The expression above corrects a minor typo in \cite{sam-weyman}[section 3.2].
} \cite{coskun}, \cite{sam-weyman}[section 3.2]
\begin{equation}
\frac{k (2n-3k-1) }{2}.
\end{equation}
For later use, note that in the special case $n=2k$,
the complex dimension of $OG^{\pm}(k,2k)$ is $(1/2) k (k-1)$.

In the special case of $OG(m,2m)$, the orthogonal Grassmannian
decomposes into a disjoint union of two spaces, denoted
$OG^{\pm}(m,2m)$:
\begin{equation}
OG(m,2m) \: = \: OG^+(m,2m) \coprod OG^-(m,2m).
\end{equation}
(This corresponds to Pl\"ucker coordinates being (anti-)self-dual.)
These two components $OG^{\pm}(m,2m)$ are isomorphic to one another.
They are also known as spinor varieties, denoted ${\cal S}_m$:
\begin{equation}
OG^+(m,2m) \: \cong \: OG^-(m,2m) \: \cong \: {\cal S}_m.
\end{equation}
(See e.g. \cite[section 6.1]{gh} for one perspective on this splitting.)

A few examples are as follows:
\begin{eqnarray}
OG^+(2,4) & = & {\mathbb P}^1 \: = \: OG(1,3), 
\\
OG^+(3,6) & = & {\mathbb P}^3 \: = \: OG(2,5), 
\\
OG^+(4,8) & = & \mbox{quadric 6-fold}.
\end{eqnarray}
When $k=1$, $OG(k,n)$ is a quadric hypersurface in 
${\mathbb P}^{n-1}$.  This follows immediately from the GLSM.
For $k=1$, the symmetric tensor representation has only one component,
so there is only one $q$ field, of charge $-2$.  It multiplies a quadric
polynomial in the $\phi$ fields, and hence coincides with the GLSM
for the vanishing locus of that quadric polynomial in ${\mathbb P}^{n-1}$.
For example, a quadric in ${\mathbb P}^2$ is\footnote{
If we didn't projectivize, the reader will note that this is an equation
for ${\mathbb C}^2/{\mathbb Z}_2$, or explicitly
\begin{displaymath}
(\phi_0)^2 + \phi_1 \phi_{-1} \: = \: 0.
\end{displaymath}  
However, after projectivization, corresponding to gauging the $U(1)$
symmetry,
this becomes a curve, and codimension-one quotient singularities do not
exist on curves as varieties.
} ${\mathbb P}^1$, and so we recover
the standard result that $OG(1,3) = {\mathbb P}^1$.

The simplest example in which to see the decomposition of 
$OG(m,2m)$ explicitly is $OG(1,2)$.  As described above, this is a quadric
hypersurface in ${\mathbb P}^1$ 
given by $\phi_1 \phi_{-1} = 0$.  This equation is reducible, and in any
event any hypersurface in ${\mathbb P}^1$ will describe a collection of
points.  In this case, we see $OG(1,2)$ is two points, so
$OG^{\pm}(1,2)$ are each a single point.
Geometrically, we can think of this in terms of isotropic subspaces
of ${\mathbb C}^2$ with quadratic form defined by the symmetric matrix
\begin{equation}  \label{eq:og12:matrix}
\left[ \begin{array}{cc} 0 & 1 \\ 1 & 0 \end{array} \right].
\end{equation}
There are two isotropic subspaces, each one-dimensional, one generated by
$(1,0)^T$, the other by $(0,1)^T$.  Since the subspaces are unique, there
are no deformations, and so $OG^{\pm}(1,2)$ should each be a single point.

We can understand the decomposition of $OG(2,4)$ similarly:  think of this
in terms of isotropic subspaces of two copies of the vector space above,
${\mathbb C}^4$ with a quadratic form defined by two copies of
the matrix~(\ref{eq:og12:matrix}) along the diagonal.  Then, for example,
the vectors $(1,0,1,0)^T$, $(0,1,0,1)^T$ each lie in distinct isotropic
subspaces.  In this case, however, these vectors lie in larger families,
which we can visualize by moving the two choices of ${\mathbb C}^2$'s
inside ${\mathbb C}^4$.  Such a choice is equivalent to choosing
a one-dimensional subspace of ${\mathbb C}^2$, which is ${\mathbb P}^1$,
hence we see $OG^{\pm}(2,4) = {\mathbb P}^1$.

The Euler characteristic of $OG(n,2n)$ is \cite{yz}
given by $2^n$, the same as $LG(n,2n)$.
The Euler characteristic of either chiral component
$OG^{\pm}(n,2n)$ is $2^{n-1}$.

Global symmetries of this QFT follow the same pattern discussed
earlier.  We can rotate the chiral superfields into one another,
preserving the superpotential.  For $OG(k,n)$, we have
$n$ chiral superfields, and chiral superfield rotations are 
described by $U(n)/U(1) = PSU(n)$; restricting to those preserving
the superpotential -- in particular, those preserving the metric --
restrict to $SO(n, {\mathbb C})$ (up to possible finite group quotients).

Just as in symplectic Grassmannians, orthogonal Grassmannians $OG(k,n)$
can be trivially embedded into $G(k,n)$, and share the Pl\"ucker map
defined by $SU(k)$-invariant baryons
\begin{equation}
B_{\alpha_1 \cdots \alpha_k} \: = \:
\epsilon_{a_1 \cdots a_k} \phi^{a_1}_{\alpha_1} \cdots \phi^{a_k}_{\alpha_k},
\end{equation}
for $\alpha = \pm i$.
In the special case that $n$ is even and $k=n/2$, there is a second class
of $SU(k)$-invariant operators, just as for Lagrangian Grassmannians,
given by
\begin{equation}
\tilde{B}_{\pm 1, \cdots, \pm n} \: = \:
\epsilon_{a_1 \cdots a_n} \phi^{a_1}_{\pm 1} \cdots \phi^{a_n}_{\pm n}.
\end{equation}  
As written, these define a map from $OG(m,2m)$ into
a projective space of dimension $2^m - 1$; however, the two components
$OG^{\pm}(m,2m)$ naturally live within subsets defined by the `chiral' spinors,
and so we also have maps on the chiral components $OG^{\pm}(m,2m)$,
mapping them into projective spaces of dimension $2^{m-1} - 1$.
(See \cite[section 3.7]{gm} for more information.)  These maps are sometimes
known as the chiral spinor embeddings.

Further background on pure spinors, as relevant to these chiral
spinor embeddings, can be found in
\cite[chapters V, VI]{cartan}.
Background on spinor varieties and orthogonal Grassmannians can be
found in {\it e.g.} \cite{kt,rincon,manivel1,manivel2,ottaviani,hos-tak,lands-man,iliev-mark,ran1,muk-curves,coskun,sam-weyman},
\cite[section 6.2]{araujo}.
A GLSM describing a degree 12 K3 surface, which is a subvariety of
$OG^+(5,10)$, is example $SSSM_{1,8,5}$ in \cite[section 2.4]{Gerhardus:2015sla}.

\subsection{Mixed Higgs-Coulomb phases at $r \ll 0$}

In this section we will study the $r \ll 0$ phases of the GLSMs for
orthogonal Grassmannians.  
Here,
the $r \ll 0$ phases will be mixed Higgs-Coulomb branches,
containing both Higgs and Coulomb vacua.  (In appendix~\ref{app:mixedphases} 
we discuss
such phases in the simpler context of hypersurfaces in projective
spaces.)

One would be tempted to try to analyze the resulting theories
using the methods of Hori-Tong \cite{Hori:2006dk}.
There, one had $U(k)$ gauge theories with fundamentals as well as
fields charged only under $\det U(k)$.  These theories were analyzed
in a Born-Oppenheimer approximation, `fibering' the $SU(k)$ gauge theories
over the space of vacua defined by the fields charged only under
$\det U(k)$.  Here, however, there are no fields charged solely under
$\det U(k)$; all the fields are charged nontrivially under $SU(k)$,
so no analogous Born-Oppenheimer analysis is pertinent.

\noindent \underline{$OG(n,2n)$}

As described above, the orthogonal Grassmannian $OG(n,2n)$
is described as a GLSM with a $U(n)$ gauge group with matter
\begin{itemize}
        \item $2n$ chiral fields $\phi_i^a$ in the fundamental representation $V$,
        \item $1$ chiral field $q_{ab}$ in the symmetric tensor product representation ${\rm Sym}^2V^*$,
\end{itemize}
where $a = 1, \cdots, n$, $i = \pm 1, \cdots, \pm n$,
and superpotential
\begin{displaymath}
W = \sum_{i=1}^{n} q_{ab}\phi_i^a \phi_{-i}^b.
\end{displaymath}

Next, we consider the phase $r \ll 0$.  Here, from the $D$-terms,
the $q_{ab}$ cannot all vanish simultaneously, 
while the $F$-terms imply that the vevs of all $\phi_i^a$ should vanish. 
In particular, 
there can be a nontrivial Higgs branch when $r \ll 0$, which must be
taken into account when computing vacua.
This Landau-Ginzburg phase will play a role in the next analysis.

In general, since the $r \gg 0$ phase is Fano, one would ordinarily
expect that the $r \ll 0$ phase is accompanied by discrete Coulomb
vacua \cite{Melnikov:2006kb}.  However, in these theories, 
describing $OG(n,2n)$ specifically, there are no discrete
Coulomb vacua for $n>1$, as we will establish next.

The one-loop corrected twisted superpotential is
\begin{align*}
        \widetilde{W}_{\rm eff} =& -t\sum_{a=1}^n \Sigma_a - 
\sum_{i,a} \rho_{ia}^b \Sigma_b \left(\ln \rho_{ia}^c\Sigma_c -1\right) -
         \sum_{\mu \geq \nu=1}^n \sum_{a=1}^n \rho^a_{\mu\nu} \Sigma_a \left[\ln\left( \sum_{b=1}^n\rho^b_{\mu\nu} \Sigma_b \right) - 1 \right] \\
         &\quad - \sum_{\mu,\nu = 1}^n\sum_{a=1}^n\alpha_{\mu\nu}^a \Sigma_a
 \left[ \ln\left(\sum_{b=1}^n\alpha_{\mu\nu}^b\Sigma_b\right) - 1 \right] 
\end{align*}
with $\rho_{ia}^b =  \delta_a^b$, 
$\rho^a_{\mu\nu} =- \delta^a_\mu - \delta^a_\nu$ and 
$\alpha_{\mu\nu}^a = -\delta_\mu^a + \delta_\nu^a$, and where we have
assumed $n>1$.
Simplifying, we get
\begin{equation}
        \widetilde{W}_{\rm eff} = - (t+i (n-1)\pi )\sum_{a=1}^n \Sigma_a - \sum_{a=1}^n 2n \Sigma_a\left(\ln \Sigma_a -1\right) -
         \sum_{\mu \geq \nu=1}^n \sum_{a=1}^n \rho^a_{\mu\nu} \Sigma_a \left[\ln\left(  \sum_{b=1}^n\rho^b_{\mu\nu} \Sigma_b \right) - 1 \right], \nonumber
\end{equation}
so that 
\begin{equation}
        \frac{\partial \widetilde{W}_{\rm eff}}{\partial \sigma_a}  = - t - i (n-1)\pi  - 2n \ln \sigma_a + 2 \ln(-2\sigma_a) + \sum_{b \neq a} \ln(- \sigma_a - \sigma_b),
\end{equation}
which gives the chiral ring relations 
\begin{equation} \label{eq:ogn2n:1}
        4 q \prod_{b \neq a}(\sigma_a+\sigma_b) \: = \: \sigma_a^{2n-2},
\end{equation}
with $q = \exp(-t)$. 
However, this has no solutions that are not contained inside the
excluded locus $\{ \sigma_a \neq \sigma_b \}$.
We can see this as follows.

Our analysis follows the same form as in section~\ref{sec:lag}.  Suppose
for the moment that $n=2k+1$.  We write
equation~(\ref{eq:ogn2n:1}) as
\begin{equation}
4 q \left( (\sigma_a)^{2k} \: + \: \cdots \: + \: 
\sigma_a^2 e_{2k-2}(\sigma) \: + \: e_{2k}(\sigma) \right) \: = \:
(\sigma_a)^{2k},
\end{equation}
where the $e_i$ are elementary symmetric polynomials in all of the $\sigma_a$.
Broadly speaking, this equation should have $2k$ roots for the value of
$(\sigma_a)^2$, but since $n=2k+1$, there are $2k+1$ different values of
$\sigma_a$ that must be assigned.  (In particular, the excluded locus
condition requires that the $\sigma_a$ must all be distinct.) 
There are two possible ways to assign values.
\begin{itemize}
\item One option is if one $\sigma_a = 0$, and the others are distinct
and nonzero.  In this case, the Coulomb branch relation~(\ref{eq:ogn2n:1})
reduces to
\begin{equation}
\prod_{a \neq b} \sigma_a \: = \: 0,
\end{equation}
so at least one other value of $\sigma_a$ must vanish, giving $\sigma_a = 
\sigma_b (=0)$ for some $a \neq b$, a contradiction.
\item Another option is if two values of $\sigma$ differ only by a sign.
Suppose, without loss of generality, that $\sigma_1 = - \sigma_2$,
so that $\sigma_1^2 = \sigma_2^2$ but $\sigma_1 \neq \sigma_2$.
Then, in this case, equation~(\ref{eq:ogn2n:1}) implies
\begin{equation}
\sigma_1^{2n-2} \: = \: 0, \: \: \:
\sigma_2^{2n-2} \: = \: 0,
\end{equation}
hence $\sigma_1 = \sigma_2 = 0$, again contradicting the assumption that
the values of $\sigma$ are distinct.
\end{itemize}
The analysis for the case of $n$ even is nearly identical.
As a result, we see that for $n>1$
there are not enough distinct solutions to the
Coulomb branch relations to satisfy the excluded locus condition, 
hence there is no Coulomb branch, the $r \ll 0$ phase is pure Higgs.

Now, let us consider the Higgs branch in the phase $r \ll 0$.
In the case $n=1$, the representation defining $q$ is one-dimensional,
so the GLSM is that for a quadric hypersurface in ${\mathbb P}^1$,
with the $r \ll 0$ phase 
a ${\mathbb Z}_2$ orbifold of a Landau-Ginzburg model with superpotential
\begin{equation}
W \: = \: \phi_1 \phi_{-1},
\end{equation}
i.e. a mass term.  Since there is an even number of massive fields,
there are two vacua 
\cite{Hellerman:2006zs,Caldararu:2007tc,Hori:2011pd,Wong:2017cqs}, 
\cite[section 4.2]{Sharpe:2019ddn}.
The $r \gg 0$ phase is $OG(1,2)$, which is two points, so we see that the
Euler characteristics match, trivially.

For $r \ll 0$ for more general cases, 
$D$ terms imply that the $q_{ab}$ are not all zero,
which Higgses the gauge group $U(n)$ to ${\mathbb Z}_2 \times SO(n)$
(see e.g. \cite{haber}), and gives a mass to the $\phi$ fields.
The ${\mathbb Z}_2$ subgroup arises as a subgroup of $\det U(n)$
that acts
trivially on the $q_{ab}$, and so, from decomposition
\cite{Hellerman:2006zs,Caldararu:2007tc,Hori:2011pd,Wong:2017cqs,Sharpe:2014tca,Sharpe:2019ddn}, 
we expect that the Landau-Ginzburg geometry is a disjoint
union of two spaces, just as the $r \gg 0$ geometry.
We shall not pursue the geometry of this Landau-Ginzburg phase further in
this paper, but it would be interesting to do so, especially to compare
to the predictions for this phase from
homological projective duality \cite{hpd,kuz1,kp1,kp2}.

\noindent \underline{$OG(n,2n+1)$}

First, we shall look for discrete Coulomb vacua.  We shall find that,
unlike the case of $OG(n,2n)$, this theory does have nontrivial
discrete Coulomb vacua, as well as a nontrivial Landau-Ginzburg model.

The effective twisted superpotential is 
\begin{equation}
        \widetilde{W}_{\rm eff} = - (t+i (n-1)\pi )\sum_{a=1}^n \Sigma_a - \sum_{a=1}^n (2n+1) \Sigma_a\left(\ln \Sigma_a -1\right) -
         \sum_{\mu \geq \nu=1}^n \sum_{a=1}^n \rho^a_{\mu\nu} \Sigma_a \left[\ln\left(  \sum_{b=1}^n\rho^b_{\mu\nu} \Sigma_b \right) - 1 \right], 
\end{equation}
from which one derives the chiral ring relations 
\begin{equation}
        4 q \prod_{b \neq a}\left( \sigma_a + \sigma_b \right) = \sigma_a^{2n-1}.
\end{equation}
We will see in examples that this admits nontrivial solutions.

Let us first consider $OG(1,3)$. 
Specializing our previous discussion, the corresponding GLSM 
is a $U(1)$ gauge theory with one $q$ field of charge $-2$
and three $\phi$ fields $\phi_{0,\pm 1}$ of charge $+1$, 
with superpotential
\begin{equation}
W = q (\phi_0 \phi_0 + \phi_{-1} \phi_{1}).
\end{equation}
As discussed earlier, for $r \gg 0$, this describes
${\mathbb P}^2[2] = {\mathbb P}^1$.

Now, let us turn to the $r \ll 0$ phase.
This is a ${\mathbb Z}_2$ orbifold of a Landau-Ginzburg model with
superpotential
\begin{equation}
W \: = \: \phi_0 \phi_0 + \phi_{-1} \phi_{1},
\end{equation}
describing three massive fields.  Since there is an odd number of
massive fields in this ${\mathbb Z}_2$ orbifold, there is a single
vacuum \cite{Hellerman:2006zs,Caldararu:2007tc,Hori:2011pd,Wong:2017cqs},
\cite[section 4.2]{Sharpe:2019ddn}.

However, we also need to take into account
discrete Coulomb vacua \cite{Melnikov:2006kb}.
The chiral ring relation is
\begin{equation}
        4 q = \sigma,
\end{equation}
so we see we have one discrete Coulomb vacuum, for a total of
two vacua, matching the Euler characteristic of the large-radius
phase $OG(1,3) = {\mathbb P}^1$, as expected.

Unfortunately, since the vacua live on a combination of Coulomb and
Higgs vacua, we do not know of a method to directly compute the
product relations, as we have done previously for theories in
which all of the vacua arise on a Coulomb branch.

\subsection{Calabi-Yau condition}

As another consistency test, we briefly mention Calabi-Yau conditions.
Mathematically, the intersection of the Pl\"ucker embedding of
$OG(k,n)$ with a hypersurface of degree $n-k-1$ is Calabi-Yau.
We reproduce the same condition physically as the condition for the
sum of the charges under any $U(1)$ subgroup of the gauge group to
vanish.

The GLSM for $OG(k,n)$ is a $U(k)$ gauge theory with $n$ chirals
in the fundamental $V$, and one chiral in Sym$^2 V^*$.
Under any $U(1) \subset U(k)$, the $n$ chirals in the fundamental
contribute a total of $n$ to the sum of the $U(1)$ charges,
and the one chiral in Sym$^2 V^*$ contributes $-2 - (k-1) = -k-1$,
so that the sum of the $U(1)$ charges is
\begin{equation}
n-k-1.
\end{equation}
Under the same $U(1)$, any element of the Pl\"ucker embedding
\begin{equation}
\epsilon_{a_1 \cdots a_k} \phi^{a_1}_{i_1} \cdots
\phi^{a_k}_{i_k}
\end{equation}
has charge $1$, so we see that intersecting the image of $OG(k,n)$
with a hypersurface of degree $n-k-1$ should be Calabi-Yau, reproducing
the mathematics result.

Let us consider a set of special cases to explicitly check this
result.  Recall $OG(1,n) = {\mathbb P}^{n-1}[2]$.  The Calabi-Yau condition is
that an additional hypersurface should have degee $n-2$, which 
matches $n-k-1$.

\subsection{Orthogonal flag manifolds}

In addition to orthogonal Grassmannians, there also exist
orthogonal flag manifolds.  At the level of group cosets,
these are of the form $SO(n,{\mathbb C})/P$ for suitable parabolic
subgroups $P$.  We can describe them as submanifolds of ordinary
flag manifolds $F(k_1, \cdots, k_p, 2n)$ satisfying an isotropy condition
on the maximal vector space.

Let us briefly describe GLSMs for these flag manifolds.
We begin with the GLSM for an ordinary flag manifold
$F(k_1, \cdots, k_p, 2n)$ \cite{Donagi:2007hi}.
This is a
\begin{equation}
U(k_1) \times U(k_2) \times \cdots \times U(k_p)
\end{equation}
gauge theory with bifundamentals $({\bf k_1}, {\bf \overline{k_2}})$,
$({\bf k_2}, {\bf \overline{k_3}})$, and so forth to
$({\bf k_{p-1}}, {\bf \overline{k_{p}}})$, along with $2n$ chirals in
representation ${\bf k_p}$ of $U(k_p)$, following \cite{Donagi:2007hi}.
To build the GLSM for an orthogonal flag manifold $OF(k_1, \cdots, k_p, 2n)$,
we add a chiral superfield $q_{ab}$ transforming in
the Sym$^2 {\bf \overline{k_p}}$ representation of $U(k_p)$,
along with a superpotential
of the form
\begin{equation}
W \: = \: \sum_{ab} q_{ab} \left( \Phi_0^a \Phi_0^b \: + \:
\sum_{i=1}^m \Phi_i^a \Phi^b_{-i} \right)
\: \: \: {\rm or} \: \: \:
\sum_{ab} q_{ab} \left( \sum_{i=1}^m \Phi_i^a \Phi^b_{-i} \right)
\end{equation}
(depending upon whether $n$ is even or odd).  As for symplectic flag
manifolds, we only impose an isotropy condition on the last, maximal,
flag, as all other vector spaces in the flag are subspaces.

We will not compute quantum cohomology rings from the GLSM here,
but mathematical discussions of quantum cohomology rings for
orthogonal flag manifolds can be found in
\cite{kim}.

\subsection{Mirrors of orthogonal Grassmannians}

Now let us consider the mirror model to the above orthogonal Grassmannian. 
We will follow the nonabelian mirror ansatz discussed in
\cite{Gu:2018fpm}.  (It should be noted that other notions of
mirrors exist, with different UV presentations but apparently equivalent
IR physics, see \cite{r08,mr,rw,pr,spacek,kalashnikov}.)

\noindent \underline{$OG(k,2n)$}

According to \cite{Gu:2018fpm}, the mirror model to $OG(k,2n)$ is a Landau-Ginzburg model with 
\begin{itemize}
	\item chiral superfields $Y_{ia}$, $i\in \{ \pm 1,\cdots, \pm n \}$ and $a\in\{1,\cdots,k \}$,
	\item chiral superfields $U_{\mu\nu} = \exp(-V_{\mu\nu})$, mirror to $q_{\mu\nu}$,
	\item chiral superfields $X_{\mu\nu} = \exp(-Z_{\mu\nu})$, mirror to W-bosons,
	\item $\sigma_a$.
\end{itemize}
and the superpotential is
\begin{align*}
	W = &\sum_a \sigma_a \left( \sum_i Y_{ia} - \sum_{\mu\geq\nu} \rho^a_{\mu\nu} \ln U_{\mu\nu} - \sum_{\mu\neq\nu} \alpha^a_{\mu\nu} \ln X_{\mu\nu} - t \right) \\
	&\quad + \sum_{i,a}\exp(-Y_{ia}) + \sum_{\mu \geq \nu}U_{\mu\nu} + \sum_{\mu\neq \nu} X_{\mu\nu},
\end{align*}
where $\rho_{\mu\nu}^a = -\delta^a_\mu - \delta^a_{\nu}$. From the definition of this mirror Landau-Ginzburg model, the dimension can be counted as $2nk - k(k-1) - \frac{1}{2}k(k+1) - k = \frac{1}{2}k(4n - 3k -1)$, which matches the dimension of $OG(k,2n)$.

Now let us compute the chiral ring relation. First, integrate out $\sigma_a$'s and we will get 
\begin{equation}
	\sum_{i} Y_{ia} - \sum_{\mu\geq\nu} \rho^a_{\mu\nu} \ln U_{\mu\nu} - \sum_{\mu\neq\nu} \alpha^a_{\mu\nu} \ln X_{\mu\nu} = t,
\end{equation}
namely, we have
\begin{align*}
	Y_{na} &= t - \sum_{i<n} Y_{ia} + \sum_{\mu\geq\nu} \rho^a_{\mu\nu} \ln U_{\mu\nu} + \sum_{\mu\neq\nu} \alpha^a_{\mu\nu} \ln X_{\mu\nu}\\
	&= t - \sum_{i<n} Y_{ia} - 2\ln U_{aa} - \sum_{\mu\neq a}\ln U_{a\mu} - \sum_{\mu\neq a} \left(\ln X_{a\mu} - \ln X_{\mu a} \right).
\end{align*}
Define
\begin{equation}
\label{eq:Pia}
	\Pi_a = \exp(-Y_{na}) = q \left(\prod_{{i<n}}\exp(Y_{ia})\right) U_{aa}^2\left(\prod_{\mu \neq a}U_{a\mu}\right)\left(\prod_{\mu\neq a} \frac{X_{a\mu}}{X_{\mu a}}\right),
\end{equation}
therefore the superpotential becomes
\begin{equation}
	W = \sum_{i<n,a}\exp(-Y_{ia}) + \sum_a \Pi_a + \sum_{\mu\geq\nu}U_{\mu\nu} + \sum_{\mu\neq \nu} X_{\mu\nu}.
\end{equation}
Now let us look at the critical locus defined by 
\begin{displaymath}
\exp\left( \frac{\partial W}{\partial \phi} \right) = 1,\quad 
\text{for}\ \phi\ \text{an arbitrary field.}
\end{displaymath}
In components, for each $a$, we have
\begin{align*}
	\frac{\partial W}{\partial Y_{ia}} &= - \exp(-Y_{ia}) + \Pi_a, \quad \text{for}\ i < n, \\
	\frac{\partial W}{\partial U_{aa}} &= 1 + \frac{2}{U_{aa}}\Pi_a, \\
	\frac{\partial W}{\partial U_{a\mu}} &= 1 + \frac{1}{U_{a\mu}} \Pi_a + \frac{1}{U_{a\mu}} \Pi_\mu,\quad \text{for}\ \mu\neq a,\\
	\frac{\partial W}{\partial X_{a\mu}} &= 1 + \frac{1}{X_{a\mu}} \Pi_a - \frac{1}{X_{a\mu}} \Pi_\mu, \quad \text{for}\ \mu\neq a, \\
	\frac{\partial W}{\partial X_{\mu a}} &= 1 - \frac{1}{X_{\mu a}} \Pi_a + \frac{1}{X_{\mu a}} \Pi_\mu,\quad \text{for}\ \mu\neq a,
\end{align*}
where in the third equation, we have used $U_{a\mu} = U_{\mu a}$. 
Therefore, on the critical locus, we have
\begin{equation}
	 \exp(-Y_{ia})  = \Pi_a, \quad - U_{a\mu} = \Pi_a + \Pi_\mu ,\quad X_{\mu \nu} = - \Pi_\mu + \Pi_\nu. \nonumber
\end{equation}
Then plugging back into equation~(\ref{eq:Pia}), we have
\begin{equation}
	\Pi_a^{2n-2} = 4 q \prod_{\mu \neq a} (\Pi_a  + \Pi_\mu).
\end{equation}
The chiral ring relations obtained from mirror models are equivalent to each other given that
$$\Pi_a \leftrightarrow \sigma_a . $$

\noindent \underline{$OG(k,2n+1)$}

The mirror to $OG(k,2n+1)$ is defined as the Landau-Ginzburg model with
\begin{itemize}
	\item chiral superfields $Y_{ia}$, $i\in \{ 0, \pm 1,\cdots, \pm n \}$ and $a\in\{1,\cdots, k \}$,
	\item chiral superfields $U_{\mu\nu} = \exp(-V_{\mu\nu})$, mirror to $q_{\mu\nu}$,
	\item chiral superfields $X_{\mu\nu} = \exp(-Z_{\mu\nu})$, mirror to W-bosons,
	\item $\sigma_a$,
\end{itemize}
and the superpotential is
\begin{align*}
	W = &\sum_a \sigma_a \left( \sum_i Y_{ia} - \sum_{\mu\geq\nu} \rho^a_{\mu\nu} \ln U_{\mu\nu} - \sum_{\mu\neq\nu} \alpha^a_{\mu\nu} \ln X_{\mu\nu} - t \right) \\
	&\quad + \sum_{i,a}\exp(-Y_{ia}) + \sum_{\mu \geq \nu}U_{\mu\nu} + \sum_{\mu\neq \nu} X_{\mu\nu},
\end{align*}
where $\rho_{\mu\nu}^a = -\delta^a_\mu - \delta^a_{\nu}$. From the definition of this mirror Landau-Ginzburg model, the dimension can be counted as $(2n+1)k - k(k-1) - \frac{1}{2}k(k+1) - k = \frac{1}{2}k(4n-3k+1)$, which matches the dimension of $OG(k,2n+1)$. 

First integrate out $\sigma_a$'s,
\begin{align}
	Y_{0a} &= t - \sum_{i\neq 0} Y_{ia} + \sum_{\mu\geq\nu} \rho^a_{\mu\nu} \ln U_{\mu\nu} + \sum_{\mu\neq\nu} \alpha^a_{\mu\nu} \ln X_{\mu\nu}, \nonumber \\
	&= t - \sum_{i\neq 0} Y_{ia} - 2 \ln U_{aa} - \sum_{\mu \neq a} \ln U_{a\mu} - \sum_{\mu \neq a }\left( \ln X_{a\mu} - \ln X_{\mu a} \right), 
\end{align}
and define
\begin{equation}
	\Pi_a = \exp\left( - Y_{0a} \right) = q \left(\prod_{{i \neq 0}}\exp(Y_{ia})\right) U_{aa}^2\left(\prod_{\mu \neq a}U_{a\mu}\right)\left(\prod_{\mu\neq a} \frac{X_{a\mu}}{X_{\mu a}}\right).
\end{equation}
The superpotential becomes
\begin{equation}
	W = \sum_{i\neq 0,a}\exp(-Y_{ia}) + \sum_a \Pi_a + \sum_{\mu\geq\nu}U_{\mu\nu} + \sum_{\mu\neq \nu} X_{\mu\nu}.
\end{equation}
Using the same calculations as in case of $OG(k,2n)$, we have
\begin{equation}
	\exp(-Y_{ia})  = \Pi_a, \quad - U_{a\mu} = \Pi_a + \Pi_\mu ,\quad X_{\mu \nu} = - \Pi_\mu + \Pi_\nu. \nonumber
\end{equation}
and 
\begin{equation}
	\Pi_a^{2n-1} = 4q \prod_{\mu \neq a} \left( \Pi_a + \Pi_\mu \right),
\end{equation}
which is the same as the chiral ring relations for $OG(k,2n+1)$ by
$$\Pi_a \leftrightarrow \sigma_a.$$

\section{Conclusions}

In this paper we have studied GLSM realizations of symplectic and orthogonal
Grassmannians and flag manifolds, 
which is to say, spaces of the form $SO(n)/P$ and
$Sp(n)/P$ for suitable subgroups $P$, generalizing GLSMs for ordinary
Grassmannians $G(k,n)$ which are of the form $U(n)/P$.  We have checked
our descriptions by comparing ordinary and equivariant quantum cohomology
rings predicted by GLSMs with those derived mathematically, compared
Witten indices of different phases.  We have also discussed mirrors of
these GLSMs.

One future direction is to generalize to GLSMs for
Grassmannians and flag manifolds
derived from exceptional groups.  Another direction is to understand
how to interpret the various phases in terms of homological projective
duality \cite{hpd}.

\section{Acknowledgements}

We would like to thank N.~Addington, R.~Donagi, R.~Eager, W.~Li,
G.~Lockhart, K.~Xu, Y.~Zhou and especially
L.~Mihalcea for
useful conversations, and D.~Bykov for a close reading of a draft.
E.S. was partially supported by NSF grant PHY-1720321.

\appendix

\section{Symmetric polynomials}
\label{app:symm}

In this section, we briefly define two classes of symmetric polynomials
and list some identities which are used
extensively in this paper.

\noindent \underline{Elementary symmetric polynomials}

The $k$th elementary symmetric polynomial in $n$ variables $x_1, \cdots, x_n$,
denoted $e_k(x)$, is defined by
\begin{equation}
        e_k(x) = \sum_{1\leq i_1 < \cdots < i_k \leq n} x_{i_1} x_{i_2} \cdots x_{i_k}.
\end{equation}
with $e_0(x) =1 $ and $e_k(x) = 0$ for $k<0$.
For example, the elementary symmetric polynomials in $3$ variables $x_1$, $x_2$ and $x_3$ include
\begin{align*}
        & e_1(x) = x_1 + x_2 + x_3, \\
        & e_2(x) = x_1 x_2 + x_1 x_3 + x_2 x_3, \\
        & e_3(x) = x_1 x_2 x_3.
\end{align*}

It can be shown that for elementary symmetric polynomials in
$n$ variables $x_1, \cdots, x_n$,
\begin{equation}  \label{eq:app:symm:e1}
        e_{\ell}(x)^2 + 2 \sum_{j=1}^{n-l} (-1)^j e_{\ell+j}(x) e_{\ell-j}(x) = \sum_{1 \leq i_1 < \dots < i_{\ell} \leq n}x_{i_1}^2\dots x_{i_{\ell}}^2.
\end{equation}

\noindent \underline{Complete homogeneous symmetric polynomials}

The $k$-th complete homogeneous symmetric polynomial in $n$ variables
$x_1, \cdots, x_n$, denoted $h_k(x)$, is defined by
\begin{equation}
        h_k(x) = \sum_{1\leq i_1 \leq \cdots \leq i_k \leq n} x_{i_1} x_{i_2} \cdots x_{i_k},
\end{equation}
with $h_0(x) =1 $ and $h_k(x) = 0$ for $k<0$.
They can also be defined as
\begin{equation}
        h_k(x) = \sum_{i_1 + \cdots + i_n = k} x_1^{i_1} x_2^{i_2} \cdots x_n^{i_n},
\end{equation}
where $i_1, \dots, i_n$ are non-negative integers.
For example, the complete homogeneous symmetric
polynomials in $3$ variables $x_1$, $x_2$ and $x_3$ include
\begin{align*}
        & h_1(x) = x_1 + x_2 + x_3, \\
        & h_2(x) = x_1^2 + x_2^2 + x_3^2 + x_1 x_2 + x_1 x_3 + x_2 x_3, \\
        & h_3(x) = x_1^3 + x_2^3 + x_3^3 + x_1^2 x_2 +x_1^2 x_3 + x_1 x_2^2 +x_2^2 x_3 + x_1 x_3^2 + x_2 x_3^2 +x_1 x_2 x_3.
\end{align*}

For complete homogeneous symmetric polynomials in two variables $x_{1,2}$,
define
\begin{align*}
        P_1^{(n)} &= h_{n}(x)^2 + 2 \sum_{i=1}^{n-2} (-)^i h_{n-i}(x) h_{n+i}(x).\\
        P_2^{(n)} &= h_{n}(x)^2 + 2 \sum_{i=1}^{n} (-)^i h_{n+i}(x)h_{n-i}(x),
\end{align*}
for a given integer $n \geq 2$.  We will use the following identity in 
section~\ref{sec:gsp}:
\begin{equation}
\label{eq:cspidentity}
         2 P_1^{(n)} + 3\left(x_1^2+x_2^2\right) P_2^{(n-1)} + 4 x_1 x_2 P_2^{(n-1)}  = (-)^{n-1} \left(x_1^{2n} + x_2^{2n} \right).
\end{equation}

As we do not know a reference where this is written explicitly,
we briefly outline an argument for this identity here.
First, it is straightforward to check that it is true for $n \leq 3$,
so we will use induction to argue it for general $n$.
Assume it is true for $n$, $n+1$, then we will argue it is true for $n+2$.
Now, for polynomials in two indeterminates,
\begin{displaymath}
        h_n(x) \: = \: h_{n-1}(x) e_1(x) - h_{n-2}(x) e_2(x),
\end{displaymath}
which one can use to show
\begin{eqnarray*}
 P_1^{(n+2)}  & = &
 P_1^{(n+1)} e_1^2  + P_1^{(n)} e_2^2  + 2 (-)^n e_1^3 h_{2n+1} - 2 (-)^n e_1^2 e_2 h_{2n}   +  2 (-)^n e_2^2 h_{2n},
\\
 P_2^{(n+1)} & = &
 P_2^{(n)} e_1^2  + P_2^{(n-1)} e_2^2 - 2 (-)^n e_1 h_{2n+1} + 2 (-)^n e_2 h_{2n},
\end{eqnarray*}
where we have used $e_1 = h_1$.
This implies
\begin{eqnarray*}
\lefteqn{
         2 P_1^{(n+2)} + \left(3x_1^2+3x_2^2+4 x_1 x_2\right) P_2^{(n+1)}
} \\
        &=& e_1(x)^2 (-)^{n}\left(x_1^{2n+2} + x_2^{2n+2}\right) \: + \:
e_2(x)^2 (-)^{n-1}\left(x_1^{2n} + x_2^{2n}\right)
\\
& &  \: + \:
 2 (-)^{n+1}  e_1 h_{2n+3}  \: + \: 2 (-)^n  e_1 e_2 h_{2n+1}, \\
        &=& (-)^n\left(x_1^{2n+4}+x_2^{2n+4}\right)
\: + \: 2 (-)^n \left(x_1^{2n+3} x_2 + x_1 x_2^{2n+3}\right) 
\\
& & \: + \: 2 (-)^{n+1} e_1 h_{2n+3} 
\: + \: 2 (-)^n e_1 e_2 h_{2n+1}.
\end{eqnarray*}
For the last two terms, using the second definition of complete homogeneous symmetric polynomials, we have
\begin{eqnarray*}
\lefteqn{
         2 (-)^{n+1} e_1 h_{2n+3} \: + \: 2 (-)^n e_1 e_2 h_{2n+1}
} \\
        &=& 2 (-)^{n+1}  \sum_{i_1 + i_2 = 2n+3} \left(x_1^{i_1+1} x_2^{i_2}+x_1^{i_1} x_2^{i_2+1}\right) \: + \: 2 (-)^n \sum_{j_1 + j_2 = 2n+1} \left(x_1^{j_1+2} x_2^{j_2 + 1}+x_1^{j_1+1} x_2^{j_2+2}\right), \\
        &=& 2 (-)^{n+1}  (x_1^{2n+4}+x_2^{2n+4}) \: + \: 2 (-)^{n+1}  (x_1^{2n+3} x_2 + x_1 x_2^{2n+3}), \\
        & & \quad + \: 2 (-)^{n+1}  \sum_{i_1' + i_2' = 2n+1} \left(x_1^{i_1'+2} x_2^{i_2'+1}+x_1^{i_1'+1} x_2^{i_2'+2}\right)
\\
& & \quad  + \: 2 (-)^n  \sum_{j_1 + j_2 = 2n+1} \left(x_1^{j_1+2} x_2^{j_2 + 1}+x_1^{j_1+1} x_2^{j_2+2}\right), \\
        &=& 2 (-)^{n+1} (x_1^{2n+4}+x_2^{2n+4}) \: + \: 2 (-)^{n+1}  (x_1^{2n+3} x_2 + x_1 x_2^{2n+3}) .
\end{eqnarray*}
Therefore,
\begin{equation}
        2 P_1^{(n+2)} + \left(3x_1^2+3x_2^2+4 x_1 x_2\right) P_2^{(n+1)}
\: = \: (-)^{n+1}\left(x_1^{2n+4}+x_2^{2n+4}\right),
\end{equation}
establishing the induction.

\section{Equivariant quantum cohomology}
\label{app:equiv}

Equivariant quantum cohomology can be obtained from gauged linear sigma models by turning on twisted masses for global symmetries \cite{Hori:2000kt}. 
In this section, we will review how this works in detail for
projective spaces and Grassmannians, comparing to known
math results \cite{mihalcea2008giambelli}.

Mathematically, many results on equivariant cohomology on these
spaces follow from the universal sequence over any Grassmannian
\begin{equation} \label{eq:gkn:univ}
        0 \: \longrightarrow \: {\cal S} \: \longrightarrow \: {\cal V} 
\: \longrightarrow \: {\cal Q} \: \longrightarrow \: 0,
\end{equation}
where ${\cal S}$ is the universal subbundle, ${\cal Q}$ the universal
quotient bundle, and ${\cal V}$ a trivial bundle.  For $G(k,N)$,
${\cal S}$ has rank $k$, ${\cal Q}$ has rank $N-k$, and the
trivial bundle ${\cal V}$ has rank $N$.  If we turn on equivariant
parameters with respect to the maximal torus in $GL(N)$, then we write
${\cal V}$ as a sum of eigenspaces for the action:
\begin{equation}
        {\cal V} \cong {\mathbb C}_{t_1}\oplus{\mathbb C}_{t_2}\oplus\dots\oplus{\mathbb C}_{t_N},
\end{equation}
for generic equivariant parameters $t_1, \dots, t_n$.
The total Chern class in equivariant cohomology is given by
\begin{equation}
        c({\cal V}) \: = \: (1+t_1)(1+t_2)\cdots(1+t_N).
\end{equation}
The equivariant cohomology ring of the Grassmannian can be expressed
in terms of the equivariant Chern classes of ${\cal S}$.
For later use, the resulting expressions can often be efficiently
written in terms of functions $h_N(x|t)$, known as the
factorial complete homogeneous Schur functions,
which are defined as
\begin{equation}
\label{eq:facthomoschur}
        h_p(x|t)= \sum_{1\leq i_1 \leq \dots \leq i_p \leq  k} \left(x_{i_1} - t_{i_1}\right)\left(x_{i_2} - t_{i_2+1}\right)\dots \left( x_{i_p} - t_{i_p+p-1} \right),
\end{equation}
for $p$ an integer $N-k+1\leq p \leq N$.

\subsection{Projective spaces}

From the GLSM for the projective space ${\mathbb P}^{N-1}$, 
the chiral ring relation after turning on twisted masses is given by
\begin{equation} \label{eq:eqc:proj:phys}
	\prod_{i=1}^N \left( \sigma - m_i \right) = q.
\end{equation}

Mathematically, for ${\mathbb P}^{N-1}$,
the equivariant quantum cohomology ring relation is\footnote{
We use $\tilde{q}$ to distinguish from the $q$ in ordinary chiral ring relations.}  \cite{mihalcea2008giambelli}
\begin{equation} \label{eq:eqc:proj}
        h_N(x|t) = \tilde{q}.
\end{equation}
In this case, $k=1$, $p$ can only be $N$ and $t = (t_1,\dots,t_N)$,
\begin{equation}
        h_N(x|t) = (x - t_1)(x - t_2)\dots (x - t_{N}),
\end{equation}
so the mathematical relation~(\ref{eq:eqc:proj}) for equivariant
quantum cohomology
matches the physical chiral ring
relation~(\ref{eq:eqc:proj:phys}) if we identify
$x=\sigma$, $t_i = m_i$, and $\tilde{q}=q$. 

In terms of the universal subbundle ${\cal S}$ and its equivariant
Chern classes, from from~(\ref{eq:gkn:univ}) we have
\begin{align*}
        c_1({\cal S}) + c_1({\cal Q}) &= e_1 (t),\\
        c_1({\cal S})c_1({\cal Q}) + c_2({\cal Q}) &= e_2 (t), \\
        \dots &  \\
        c_1({\cal S})c_{N-1}({\cal Q}) + c_N({\cal Q}) &= e_N (t),
\end{align*}
or more simply,
\begin{equation}
        c_{\ell}({\cal Q}) = e_{\ell}(t) - c_1({\cal S}) c_{\ell-1}({\cal Q}) = \sum_{i=0}^{\ell} (-c_1({\cal S}))^i e_{\ell-i}(t),
\end{equation} 
for $\ell = 1, 2, \dots, N$,
where $e_i(t)$ is the $i$-th elementary symmetric polynomial of $t_1,\dots,t_N$.
In particular,
\begin{equation}
        c_N({\cal Q}) = \sum_{i=0}^N (-c_1({\cal S}))^i e_{N-i}(t) = \sum_{i=0}^N (-1)^i x^i e_{N-i}(t) = (-1)^N h_N(x|t).
\end{equation}
Classically, $c_N({\cal Q}) = 0$.
In the quantum theory,
$c_N({\cal Q})=(-1)^N \tilde{q}$,
which yields the equivariant cohomology ring relations
$h_N(x|t)=\tilde{q}$ in \cite{mihalcea2008giambelli},
for $x = c_1({\cal S})$.

\subsection{Grassmannians}

Let us consider the general Grassmannians, $G(k,N)$. It can be realized in the $U(k)$ GLSM with $N$ fundamentals. The chiral ring relations are
\begin{equation}
	\prod_{i=1}^N \left( \sigma_a - m_i \right) = (-1)^{k-1} q, \quad {\rm for} \ a = 1,\dots,k.
\end{equation}

First, consider
the special case $G(2,4)$.
Define $x_1$, $x_2$ as the Chern roots of the universal subbundle
${\cal S}$.  The equivariant quantum cohomology ring relations
are given by \cite{mihalcea2008giambelli}
\begin{align*}
        & c_3({\cal Q}) \: = \: - h_3(x|t) \: = \: 0,\\
        & c_4({\cal Q}) \: =  \: h_4(x|t)  \: = \:  - \tilde{q}.
\end{align*}
We claim the GLSM predictions match.
The physical chiral ring relations for $G(2,4)$ are
\begin{align*}
	& \prod_{i=1}^4 \left( \sigma_1 - m_i \right) \: = \: - q, \\
	& \prod_{i=1}^4 \left( \sigma_2 - m_i \right) \: = \: - q.
\end{align*}
Subtracting these equations and factoring out $\sigma_1-\sigma_2$ (since
the excluded locus forbids $\sigma_1 = \sigma_2$), we have
\begin{equation}
h_3(\sigma) - e_1(m) h_2(\sigma) + e_2(m) h_1(\sigma) - e_3(m) \: = \: 0.
\end{equation}
Since the left-hand side of the equation above is $h_3(\sigma|m)$,
we recover the first 
equivariant quantum cohomology ring relation after identifying $t_i=m_i$ and 
$x_a = \sigma_a$.

The sum of the physical chiral ring relations is
\begin{equation}
(\sigma_1^4+\sigma_2^4) - (\sigma_1^3+\sigma_2^3) e_1(m) + (\sigma_1^2+\sigma_2^2) e_2(m) - (\sigma_1+\sigma_2)e_3(m) + 2 e_4(m) = -2 q.
\end{equation}
Adding $(\sigma_1+\sigma_2)h_3(\sigma|m)$ to the left-hand side 
gives $2h_4(\sigma|m)$. Therefore, we end up with
\begin{equation}
	h_4(\sigma|m) = - q,
\end{equation}
which matches the second equivariant quantum cohomology ring relation 
if we also identify $\tilde{q}=q$.

Next, consider $G(2,N)$.
The equivariant quantum cohomology ring relations are given by
\cite{mihalcea2008giambelli}
\begin{align*}
	&h_{N-1}(x|t) = 0,\\
	&h_N(x|t) = - \tilde{q}.
\end{align*}
We follow the same pattern to show that these ring relations follow
from the physical chiral ring relations
\begin{equation}
 \prod_{i=1}^N \left( \sigma_a - m_i \right) \: = \: - q,
\end{equation}
for $a\in \{1, 2\}$.
Subtracting the two equations and factoring out $\sigma_1 - \sigma_2$ yields
\begin{equation}
h_{N-1}(\sigma|m) \: = \:\sum_{i=0}^{N-1} (-1)^{i} e_{i}(m) h_{N-1-i}(\sigma) \: = \: 0,
\end{equation}
which matches the first mathematical ring relation.
Next, summing the two chiral ring relations gives
\begin{equation}
	\sum_{i=0}^{N-1}(-1)^i e_{i}(m) \left(\sigma_1^{N-i} + \sigma_2^{N-i} \right) + (-1)^N 2 e_N(m) \: = \: -2q.
\end{equation}
Since we have $h_{N-1}(\sigma|m)=0$, we can add 
$(\sigma_1+\sigma_2) h_{N-1}(m|t)$ to the left-hand side of the above equation
to get
\begin{equation}
2h_N(\sigma|m) \: = \: 
\sum_{i=0}^{N-1}(-1)^i e_{i}(m) \left(\sigma_1^{N-i} + \sigma_2^{N-i} \right) + (-1)^N 2 e_N(m) + (\sigma_1+\sigma_2) h_{N-1}(\sigma|m) \: = \: -q,
\end{equation}
from which the second mathematical relation follows.
Therefore, we see that for $\sigma_a = x_a$, $m_i = t_i$ and $q = \tilde{q}$,
the physical chiral ring relations reproduce the mathematical equivariant
cohomology relations.

Next, we consider the case $G(3,N)$.  In this case, the
equivariant quantum cohomology ring relations are \cite{mihalcea2008giambelli}
\begin{equation}
h_{N-2}(x|t) \: = \: 0,
\: \: \:
h_{N-1}(x|t) \: = \: 0,
\: \: \:
h_{N}(x|t) \: = \: q.
\end{equation}
From the GLSM for $G(3,N)$, we have following chiral ring relations:
\begin{equation}
\prod_{i=1}^N \left( \sigma_a - m_i \right) \: = \: q,
\end{equation}
for $a \in \{1, 2, 3\}$.

To derive the mathematical ring relations from the physical chiral ring
relations, we proceed as follows.
First, subtract each two of the three chiral ring relations 
and factor out $(\sigma_1-\sigma_2)$, $(\sigma_1-\sigma_3)$,
and $(\sigma_2-\sigma_3)$, to get.
\begin{align}
	&\sum_{i=0}^{N-1} (-1)^{i} e_{i}(m) h_{N-1-i}(\sigma_1,\sigma_2) \: = \: 0, \label{eq:1}\\
	&\sum_{i=0}^{N-1} (-1)^{i} e_{i}(m) h_{N-1-i}(\sigma_1,\sigma_3) \: = \: 0,\label{eq:2}\\
	&\sum_{i=0}^{N-1} (-1)^{i} e_{i}(m) h_{N-1-i}(\sigma_2,\sigma_3) \: = \: 0.\label{eq:3}
\end{align}
Subtracting any two of the equations above
gives a relation of the form
\begin{equation}  \label{eq:eqc:grass:gk3n:tmp1}
	\sum_{i=0}^{N-2} (-1)^{i} e_{i}(m) \left( h_{N-1-i}(\sigma_1,\sigma_2) - h_{N-1-i}(\sigma_1,\sigma_3) \right) \: = \: 0.
\end{equation}
We simplify this using
\begin{eqnarray*}
	h_{k}(\sigma_1,\sigma_2) - h_{k}(\sigma_1,\sigma_3) &=&
 \sum_{i+j = k} \sigma_1^i \left( \sigma_2^j - \sigma_3^j \right) 
\: = \: \left( \sigma_2 - \sigma_3 \right) \sum_{i+j = k} \sigma_1^i h_{j-1}(\sigma_2,\sigma_3), \\
	& = & \left( \sigma_2 - \sigma_3 \right) h_{k-1}(\sigma),
\end{eqnarray*}
and factoring out $\sigma_2 - \sigma_3$ from 
equation~(\ref{eq:eqc:grass:gk3n:tmp1}) then gives
\begin{equation}
\sum_{i=0}^{N-2} (-1)^{i} e_{i}(m) h_{N-2-i}(\sigma) = h_{N-2}(\sigma|m) \: = \: 0,
\end{equation}
which is the first 
equivariant quantum cohomology ring relation. 

To obtain the second ring relation, sum the three equations, 
(\ref{eq:1}), (\ref{eq:2}) and (\ref{eq:3}), which gives
\begin{eqnarray*}
0 &=& 
\sum_{i=0}^{N-1} (-1)^{i} e_{i}(m)\left( h_{N-1-i}(\sigma_1,\sigma_2) + h_{N-1-i}(\sigma_1,\sigma_3) + h_{N-1-i}(\sigma_2,\sigma_3)\right), \\
	&=& \sum_{i=0}^{N-1} (-1)^{i} e_{i}(m)\left( h_{N-1-i}(\sigma_1,\sigma_2) + h_{N-1-i}(\sigma_1,\sigma_3) + h_{N-1-i}(\sigma_2,\sigma_3)\right) \\
	&	& +  \: e_1(\sigma) h_{N-2}(\sigma|m), \\
	&=& \sum_{i=0}^{N-2} (-1)^{i} e_{i}(m)\left[ h_{N-1-i}(\sigma_1,\sigma_2) + h_{N-1-i}(\sigma_1,\sigma_3) + h_{N-1-i}(\sigma_2,\sigma_3) \right. \\
	&& \left. +  \: e_1(\sigma) h_{N-2-i}(\sigma) \right] + 3 (-1)^{N-1} e_{N-1}(m), \\
	&=& 3 \sum_{i=0}^{N-2} (-1)^{i} e_{i}(m) h_{N-1-i}(\sigma) + 3 (-1)^{N-1} e_{N-1}(m), \\
&=& 3 h_{N-1}(\sigma|m),
\end{eqnarray*}
which is the second equivariant quantum cohomology ring relation. 

The last equivariant quantum cohomology ring relation can be obtained by 
summing the three physical chiral ring relations:
\begin{eqnarray*}
	3q &	=	& \sum_{i=0}^{N-1}(-1)^i e_{i}(m) \left(\sigma_1^{N-i} + \sigma_2^{N-i} + \sigma_3^{N-i} \right) + (-1)^N 3 e_N(m), \\
&	=	& \sum_{i=0}^{N-1}(-1)^i e_{i}(m) \left(\sigma_1^{N-i} + \sigma_2^{N-i} + \sigma_3^{N-i} \right) + (-1)^N 3 e_N(m) \\
	&	& \quad + \: 2 e_1(\sigma)  h_{N-1}(\sigma|m) - e_2(\sigma) h_{N-2}(\sigma|m), \\
&	=	& 3 \sum_{i=0}^{N-1}(-1)^i e_{i}(m)h_{N-i}(\sigma) + (-1)^N 3 e_N(m), \\
&	=	& 3 h_{N}(\sigma|m).
\end{eqnarray*}
Thus, we see that all the equivariant quantum cohomology ring relations 
can be derived from the physical chiral ring relations for $G(3,N)$.

These methods are straightforward to generalize to $G(k,N)$ for $k > 3$,
and so we do not give further details here.

\section{Tensor product representation}
\label{app:product}

In this appendix, we briefly review the weights of tensor products 
of representations.
Consider first the case
$V \otimes V$, where $V$ is a two-dimensional representation. 
Denote a basis of $V$ by $\{v_1,v_2\}$, 
then a basis for the tensor product is
$$\left\{v_1\otimes v_1,v_1\otimes v_2, v_2\otimes v_1,v_2\otimes v_2\right\}.$$
In the representation $V$, the weights $\rho^a_i$
are defined by $H^a v_i = \rho_i^a v_i$, where $H^a$ is a Cartan generator \cite{georgi1999lie}. In the tensor product representation $V\otimes V$,
\begin{equation}
	H^a v_i\otimes v_j = (\rho_i^a + \rho_j^a) v_i\otimes v_j \equiv \rho^a_{ij} v_i\otimes v_j,
\end{equation}
namely, the weights are $\rho^a_{ij} = \rho_i^a + \rho_j^a$. 

We can further restrict to the (anti-)symmetric case by (anti-)symmetrizing 
the basis. In this particular example, 
the symmetric tensor product representation has the basis
\begin{displaymath}
\left\{v_1\otimes v_1,v_2\otimes v_2, \frac{1}{2} \left(v_1\otimes v_2 + v_2\otimes v_1\right)\right\},
\end{displaymath}
and the weights are $2 \rho^a_1, 2\rho^a_2, \rho^a_1+\rho^a_2 $. 
In the anti-symmetric case, the basis is
\begin{displaymath}
\frac{1}{2} \left(v_1\otimes v_2 - v_2\otimes v_1\right),
\end{displaymath}
and the weights are $\rho^a_1+\rho^a_2$. This story can be generalized easily.

\section{Equivariant quantum cohomology for \texorpdfstring{$SG(n,2n)$}{SG(n,2n)}}
\label{app:sg}

Consider the universal sequence over $SG(n,2n)$ with equivariant parameters turned on
\begin{equation}
	0 \longrightarrow {\cal S} \longrightarrow {\cal V}^{t}_{SG(n,2n)} 
 \longrightarrow {\cal Q} \longrightarrow 0.
\end{equation}
In this sequence, the tautological bundle ${\cal S}$ 
and the quotient bundle ${\cal Q}$ both have rank $n$ 
and they are dual to each other, i.e. ${\cal S}^* \cong {\cal Q}$, 
which implies $c_i({\cal S}) = (-)^i c_i({\cal Q})$ and the Chern roots of 
${\cal S}$ and ${\cal Q}$ are up to a minus sign. 
This universal bundle gives the following relation
\begin{align}
\label{eq:classicalconstraint}
	c\left({\cal V}^{t}\right)= c({\cal S})c({\cal Q}) &= (1+ c_1({\cal S}) + \cdots + c_n({\cal S}) )(1 + c_1({\cal Q}) + \cdots + c_n({\cal Q}) ),\nonumber \\
	& = (1-x_1^2)(1-x_2^2)\cdots(1 -x_n^2),
\end{align}
where $x_i$ can be the Chern roots of either ${\cal S}$ or ${\cal Q}$. The total Chern class of $c\left({\cal V}^{t}\right)$ is
\begin{equation}
\label{eq:equirvariantt}
	c({\cal V}^{t}) = (1+t_1)\cdots(1+t_n) (1-t_1) \cdots (1-t_n) = 1 - e_1(t^2) + e_2(t^2) + \cdots + (-)^n e_n(t^2).
\end{equation}
where $t_i$'s are equivariant parameters for $Sp(n)$-action and $e_i(t^2)$ is the $i$-th elementary symmetric polynomial of $\left\{t_1^2, \cdots, t_n^2\right\}$.

To obtain the (quantum) cohomology ring relations, we need to modify equation~(\ref{eq:classicalconstraint}) by adding 
$c_i({\cal S})$ and $c_i({\cal Q})$ for $n< i \leq 2n$ and it becomes
\begin{equation}
\label{eq:qringsg}
	c({\cal V}^{t}) = \left( 1 + c_1({\cal S}) + \cdots + c_{2n}({\cal S})  \right)\left( 1 + c_1({\cal Q}) + \cdots + c_{2n}({\cal Q}) \right).
\end{equation}
Since ${\cal Q} \cong {\cal S}^{*}$, 
the relations $c_i({\cal Q}) = (-)^i c_i({\cal S})$ still hold 
for $n< i \leq 2n$. 
Now the (quantum) cohomology ring relations can be obtained by extracting 
terms of the same degree from both sides. Our convention here is to choose $\left\{ c_i({\cal S})|i=1,\dots,n\right\}$ as the generators of the (quantum) cohomology and the constraints on $\left\{c_i({\cal Q})|i = n+1,\dots, 2n\right\}$ will generate the ring relations. To get the classical ones, we need to set
\begin{equation}
\label{eq:cringsgdef}
	c_{n+1}({\cal Q}) = 0,\ c_{n+2}({\cal Q}) = 0,\ \cdots,\ c_{2n}({\cal Q}) = 0.
\end{equation}
While to obtain the quantum cohomology ring relations, $\tilde{q}$ has degree $n+1$ for $SG(n,2n)$ and we need to set
\begin{equation}
\label{eq:qringsgdef}
	2 c_{n+1}({\cal Q}) = \tilde{q},\ c_{n+2}({\cal Q}) = 0,\ \cdots,\ c_{2n}({\cal Q}) = 0.
\end{equation}
In the following, we only verify the quantum case as it will 
reproduce the classical case in the limit $\tilde{q} \rightarrow 0$. It turns out we should consider two situations, $n$ odd and $n$ even, separately.

First, let us consider $n=2k+1$. Equations~(\ref{eq:equirvariantt}) and (\ref{eq:qringsg}) generate the following two sets of equations:
\begin{align*}
	& c_1({\cal S}) + c_1({\cal Q}) = 0, \\
	& c_2({\cal S}) + c_1({\cal S})c_1({\cal Q}) + c_2({\cal Q}) = - e_1(t^2), \\
	& \quad \quad \quad \cdots \\
	& c_{2k}({\cal S}) + c_{2k-1}({\cal S})c_1({\cal Q}) + \cdots + c_{1}({\cal S}) c_{2k-1}({\cal Q}) + c_{2k}({\cal Q}) = (-)^k e_k(t^2), \\
	& c_{2k+1}({\cal S}) + c_{2k}({\cal S})c_1({\cal Q}) + \cdots + c_{1}({\cal S}) c_{2k}({\cal Q}) + c_{2k+1}({\cal Q}) = 0,
\end{align*}
and 
\begin{align*}
	& c_{2k+2}({\cal S}) +  c_{2k+1}({\cal S}) c_1({\cal Q})  + \cdots + c_1({\cal S}) c_{2k+1}({\cal Q}) + c_{2k+2}({\cal Q}) = (-)^{k+1} e_{k+1}(t^2), \\
	& c_{2k+3}({\cal S}) + c_{2k+2}({\cal S}) c_1({\cal Q}) + \cdots + c_1({\cal S}) c_{2k+2}({\cal Q}) + c_{2k+3}({\cal Q}) = 0, \\
	& \quad \quad \quad \cdots \\
	& c_{4k+1}({\cal S}) + c_{4k}({\cal S}) c_{1}({\cal Q})  + \cdots + c_1({\cal S}) c_{4k}({\cal Q}) + c_{4k+1}({\cal Q}) = 0, \\
	& c_{4k+2}({\cal S}) + c_{4k+1}({\cal S}) c_{1}({\cal Q})  + \cdots + c_1({\cal S}) c_{4k+1}({\cal Q}) + c_{4k+2}({\cal Q}) = (-)^{2k+1}e_{2k+1}(t^2).
\end{align*}
Among the above two sets of relations, the relations with odd degrees are trivially satisfied due to the fact that $c_i({\cal Q}) = (-)^ic_{i}({\cal S})$. 
Therefore, we are left with $n=2k+1$ nontrivial relations of even degrees. 
In the first set, substituting $c_i({\cal Q}) = (-)^ic_{i}({\cal S})$, for $1\leq i \leq k$, we have
\begin{equation}
	2 c_{2i}({\cal S}) - 2 c_{2i-1}({\cal S}) c_{1}({\cal S}) + \cdots + (-)^{i-1} 2 c_{i+1}({\cal S}) c_{i-1}({\cal S}) + (-)^{i} c_{i}^2({\cal S}) = (-)^i e_i(t^2), \nonumber
\end{equation}
or equivalently,
\begin{equation}
	c_{i}^2({\cal S}) + 2 \sum_{l=1}^{i} (-)^{l} c_{i-l}({\cal S}) c_{i+l}({\cal S}) = e_i(t^2). \nonumber
\end{equation}
Written in terms of Chern roots of ${\cal S}$: 
\begin{equation}
\label{eq:sgqringoddset1}
	e_i(x)^2 + 2\sum_{l=1}^{i} (-)^{l} e_{i-l}(x) e_{i+l}(x) = e_i(t^2), \ {\rm for} \ i = 1, 2, \cdots, k.
\end{equation}
This set of relations will be the same for both classical case and quantum case.
For the second set, we first need to write down $c_i({\cal Q})$, 
$2k+2\leq i \leq 4k+2$, in terms of $\{c_i({\cal S})|i=1,\cdots, n = 2k+1 \}$ 
and then use equation~(\ref{eq:qringsgdef}) to obtain final results. 
For $0\leq a \leq k$, we have
\begin{align*}
	&(-1)^{k+a+1} c_{k+1+a}^2({\cal S}) + 2 (-1)^{k+a+1} \sum_{l=1}^{k-a} (-)^{l} c_{k+a+1-l}({\cal S}) c_{k+a+1+l}({\cal S}) + 2 c_{2a}({\cal S}) c_{2k+2}({\cal Q})   \\ 
	& + 2 c_{2a-1}({\cal S})c_{2k+3}({\cal Q}) + \dots + 2 c_1({\cal S})c_{2k+2a+1}({\cal Q}) + 2c_{2k+2a+2}({\cal Q})  = (-)^{k+a+1}e_{k+a+1}(t^2),
\end{align*}
and applying equation~(\ref{eq:qringsgdef}), this reduces to
\begin{equation}
	c_{k+1+a}^2({\cal S}) + 2 \sum_{l=1}^{k-a} (-)^{l} c_{k+a+1-l}({\cal S}) c_{k+a+1+l}({\cal S}) = e_{k+a+1}(t^2) + (-)^{k+a+2} c_{2a}({\cal S}) \tilde{q}. \nonumber
\end{equation}
Or equivalently, in terms of Chern roots of ${\cal S}$: 
\begin{equation}
\label{eq:sgqringoddset2}
	e_{i}^2(x) + 2 \sum_{l=1}^{2k+1-i} (-)^{l} e_{i-l}(x) e_{i+l}(x) = e_{i}(t^2) + (-)^{i+1} e_{2i-2k-2}(x) \tilde{q},\ {\rm for} \ i = k+1, \cdots, 2k+1.
\end{equation}
For the case $n=2k$, the strategy is the same and we just write down the results:
\begin{align*}
	&e_i(x)^2 + 2\sum_{l=1}^{i} (-)^{l} e_{i-l}(x) e_{i+l}(x) = e_i(t^2), &&{\rm for} \ i = 1, 2, \cdots, k. \\
	&e_{i}^2(x) + 2 \sum_{l=1}^{2k-i} (-)^{l} e_{i-l}(x) e_{i+l}(x) = e_i(t^2) + (-)^{i+1} e_{2i-2k-1}(x) \tilde{q}, && {\rm for} \ i = k+1, \cdots, 2k.
\end{align*}

We can summarize the equations for even $n$ and odd $n$ cases above in a more compact form as
\begin{equation}
\label{eq:sgqringeqiv}
	e_{i}^2(x) + 2 \sum_{l=1}^{n-i} (-)^{l} e_{i-l}(x) e_{i+l}(x) = e_{i}(t^2)+ (-)^{i+1} e_{2i-n-1}(x) \tilde{q},
\end{equation}
for $i = 1, \cdots, n$.

\section{Simple examples of mixed Higgs-Coulomb branches}
\label{app:mixedphases}

In this appendix, we will outline some simple examples of theories
with mixed Higgs-Coulomb branches, and the limitations of computing
quantum cohomology with $\sigma$ fields in each case.

First, consider the case of a hypersurface of degree $d$ in
${\mathbb P}^4$.
\begin{itemize}
\item $d=0$.  In this case, there is no hypersurface, this is just the
GLSM for ${\mathbb P}^4$ itself.  In this case, there is no Higgs branch
for $r \ll 0$, only a Coulomb branch, with $\sigma$ fields obeying
\begin{equation}
\sigma^5 \propto q.
\end{equation}
There are then five solutions for $\sigma$, matching the Euler
characteristic of ${\mathbb P}^4$, and those $\sigma$ fields can be used
to reproduce the quantum cohomology ring of ${\mathbb P}^4$, using known
methods \cite{Morrison:1994fr}.  
\item $d=5$.  This is the Calabi-Yau case.  In this case, there is no
Coulomb branch, no discrete Coulomb vacua, only the Higgs branch,
corresponding to the Landau-Ginzburg orbifold phase of this Calabi-Yau
hypersurface.  
\item $d=2$.  (This GLSM provides an alternative physical realization
of the space $SG(2,4)$, which as a variety coincides with
${\mathbb P}^4[2]$.)
This is an example of a mixed branch, with both Higgs and
Coulomb vacua.  The Coulomb vacua are solutions of
\begin{equation}
\sigma^5 = (-2 \sigma)^2 q,
\end{equation}
or $\sigma^3 \propto q$, which only has three solutions.  In addition, there
is a Landau-Ginzburg orbifold, a ${\mathbb Z}_2$ orbifold of a theory with
superpotential of the form
\begin{equation}
W \: = \: x_1^2 \: + \: \cdots \: x_5^2.
\end{equation}
Here, the $x$ fields are clearly massive, and as there is an odd number
of them, taking the ${\mathbb Z}_2$ orbifold only results in a single
vacuum \cite{Hellerman:2006zs,Caldararu:2007tc,Hori:2011pd,Wong:2017cqs},
\cite[section 4.2]{Sharpe:2019ddn}.   
Combining the Landau-Ginzburg and Coulomb vacua,
we have a total of 4 vacua, matching the Euler characteristic of 
the hypersurface ${\mathbb P}^4[2]$.  
\end{itemize}

As a related example, consider the GLSM for a hypersurface of degree $d$
in ${\mathbb P}^3$.
\begin{itemize}
\item The degree $d=0$ and $d=4$ cases follow the same form as above.
In one case, one has the GLSM for ${\mathbb P}^3$, which only has
a Coulomb branch, no Landau-Ginzburg phase.  In the other case,
one only has a Landau-Ginzburg phase, no discrete Coulomb vacua.
\item The degree $d=2$ case here is a bit more interesting.
The Coulomb vacua are solutions of
\begin{equation}
\sigma^4 \: = \: (-2\sigma)^2 q,
\end{equation}
or $\sigma^2 \propto q$, and so there are 2 discrete Coulomb vacua.
The Landau-Ginzburg phase is a ${\mathbb Z}_2$ orbifold of a theory
with superpotential of the form
\begin{equation}
W \: = \: x_1^2 \: + \: \cdots \: + \: x_4^2.
\end{equation}
Again, the $x$ fields are all massive, but there is an even number of them,
so now the ${\mathbb Z}_2$ orbifold results in two vacua 
\cite{Hellerman:2006zs,Caldararu:2007tc,Hori:2011pd,Wong:2017cqs},
\cite[section 4.2]{Sharpe:2019ddn}.  
Combining the Landau-Ginzburg vacua and Coulomb vacua, we have a total
of four vacua, which matches the Euler characteristic of
${\mathbb P}^3[2]$.  
\end{itemize}

In passing, the Coulomb branch relation for a hypersurface of degree $k$
in a projective space ${\mathbb P}^n$, namely
\begin{equation}
\sigma^{n+1} \: = \: q (-k)^k \sigma^k,
\end{equation}
appear in discussions of the quantum cohomology ring of hypersurfaces
in \cite[equ'n (16), (64)]{Jinzenji:1995kp},
\cite[equ'n (1.1)]{Collino:1996my}, as a distinguished subring of the
quantum cohomology ring (computed by the Coulomb branch of the GLSM).

\section{Dualities and examples}   \label{app:dualities}

In this appendix we will summarize some geometric relationships between various
Grassmannians, that have appeared sporadically throughout the text.

\begin{enumerate}
\item $G(k,n) \cong G(n-k,n)$,
\item $SG(1,2n) \cong {\mathbb P}^{2n-1}$,
\item $OG(1,n) \cong {\mathbb P}^{n-1}[2]$,
\item $OG(n,2n+1) \cong OG^+(n+1,2(n+1))$, see e.g. \cite[exercise 23.53]{fulton},
\item $OG(1,3) \cong OG^+(2,4) \cong {\mathbb P}^1$, 
\item $OG(1,5) \cong SG(2,4) \cong {\mathbb P}^4[2]$, see e.g.
\cite[exercise 23.50]{fulton},
\item $OG(2,5) \cong SG(1,4) \cong {\mathbb P}^3$, see e.g.
\cite[exercise 23.50]{fulton},
\item $OG(1,6) \cong G(2,4) \cong {\mathbb P}^5[2]$,
see e.g. \cite[section 23.3]{fulton},
\item $OG^+(3,6) \cong SG(1,4) \cong {\mathbb P}^3$,
see e.g. \cite[section 23.3]{fulton},
\item $OG(1,8) \cong OG^+(4,8)$, see e.g.
\cite[section 23.3]{fulton}.
\end{enumerate}

Physically, these all correspond to various IR dualities between
GLSMs, sometimes relating abelian GLSMs to nonabelian GLSMs.

\end{document}